\documentclass[aps,prb,reprint,showpacs,superscriptaddress]{revtex4-2}

\usepackage{amsmath}
\usepackage{graphicx}
\usepackage{lmodern}
\usepackage{amsmath}
\usepackage{comment}
\usepackage{color}
\usepackage{amssymb}
\usepackage{bm}
\usepackage{braket}
\usepackage{mathtools}
\usepackage[normalem]{ulem}

\newcommand{\bk}{\bm{k}}
\newcommand{\bp}{\bm{p}}
\newcommand{\bq}{\bm{q}}
\newcommand{\bQ}{\bm{Q}}

\newcommand{\ba}{\nu_c}

\usepackage[dvipsnames]{xcolor}
\usepackage[colorlinks=true]{hyperref}
\hypersetup{
    colorlinks=true,  
    linkcolor=blue,
    citecolor=blue,
    urlcolor=blue
} 

\makeatletter
\def\maketitle{
\@author@finish
\title@column\titleblock@produce
\suppressfloats[t]}
\makeatother

\begin{document}
\title{Superconductivity and magnetism in bilayer nickelates: itinerant perspective}

\author{Yi-Ming Wu}
\email{yimwu@stanford.edu}
\affiliation{Leinweber Institute for Theoretical Physics, Stanford University, Stanford, CA 94305, USA}

\author{Tobias Helbig}
\email{helbig@stanford.edu}
\affiliation{Leinweber Institute for Theoretical Physics, Stanford University, Stanford, CA 94305, USA}

\author{Salahudin V. Smailagić}
\affiliation{Leinweber Institute for Theoretical Physics, Stanford University, Stanford, CA 94305, USA}

\author{Hao-Xin Wang}
\affiliation{Department of Physics, The Chinese University of Hong Kong, Sha Tin, New Territories, Hong Kong, China}

\author{Yijun Yu}
\affiliation{Stanford Institute for Materials and Energy Sciences, SLAC National Accelerator Laboratory,
Menlo Park, CA 94025, USA}
\affiliation{Department of Applied Physics, Stanford University, Stanford, CA 94305, USA}
\affiliation{Department of Physics, Fudan University, Shanghai 200438, China}

\author{Harold Y. Hwang}
\affiliation{Stanford Institute for Materials and Energy Sciences, SLAC National Accelerator Laboratory,
Menlo Park, CA 94025, USA}
\affiliation{Department of Applied Physics, Stanford University, Stanford, CA 94305, USA}

\author{Srinivas Raghu}
\affiliation{Leinweber Institute for Theoretical Physics, Stanford University, Stanford, CA 94305, USA}

\begin{abstract}
We study superconductivity and magnetism in bilayer nickelates from an itinerant perspective. Starting from a tight binding fit to recent ARPES measurements on compressively strained thin films, we incorporate the standard set of onsite repulsive interactions among partially filled $e_g$ orbitals: intra-orbital $U$, inter-orbital $U'$, Hund's coupling $J_H$ and a pair hopping $J_P$. We obtain the effective pairing interaction by dressing these bare interactions with particle-hole fluctuations via the RPA. In the strong Hund's coupling regime, we find that $s$-wave superconductivity and $(\pi/2, \pi/2)$ SDW order are the favored ground states. With weaker Hund's coupling, we find that $d$-wave pairing and $(\pi, \pi)$ SDW are the leading ground states. Our results are qualitatively consistent with earlier DMRG studies, and point to the key role played by Hund's coupling in determining the nature of superconductivity and magnetism in this system.
\end{abstract}

\maketitle

\date{\today}
\section{Introduction} % (fold)
\label{sec:introduction}

The bilayer nickelates are a newly discovered family of high-$T_c$ superconductors that have been under intensive study in recent years. The bulk material becomes superconducting under high pressure with a transition temperature $T_c\approx80$K~\cite{Sun2023,Zhang2024exp,Hou_2023,Liu2024,Yang2024,ZHANG2024147}, and more recently, superconductivity with $T_c\approx40$K has been observed in compressively strained thin film samples~\cite{Ko2024,Zhou2025,Liu2025}. 
As a function of pressure or compressive strain, the system undergoes a structural transition from a low pressure orthorhombic~\cite{wang2024,bhatt2025resolvingstructuraloriginssuperconductivity} to nearly tetragonal system at high pressure/compressive strain.   At ambient pressure or without compressive strain, superconductivity is absent; instead, spin stripe order with {$\bm{Q}=(\pi/2,\pi/2)$} has been observed ~\cite{Chen2024,PhysRevLett.132.256503,Khasanov2025,ZHAO20251239,kakoi2024multiband,XIE20243221,Gupta2025,Ren2025}.  Such magnetic order is distinct from the conventional  $(\pi,\pi)$ Néel order observed for example in the La-based cuprate parent compound La$_2$CuO$_4$~\cite{PhysRevLett.59.1613,PhysRevB.37.7443}. 
The microscopic origin of the spin stripe phase and its connection to superconductivity 
are actively being debated at present~\cite{PhysRevB.108.L140505,PhysRevB.108.L201108,PhysRevLett.132.146002,PhysRevLett.132.036502,PhysRevB.108.174511,PhysRevB.110.104517,PhysRevLett.131.126001,PhysRevLett.132.146002,PhysRevB.108.174511,PhysRevB.110.104517,PhysRevB.108.L140505,PhysRevB.111.174506,shen2023effective,PhysRevB.111.L020504,PhysRevLett.131.206501,PhysRevLett.131.236002,PhysRevB.109.L081105,lu2023superconductivity,PhysRevLett.132.126503,PhysRevB.109.165154,PhysRevLett.133.126501,PhysRevB.108.L140504,PhysRevB.108.174501,PhysRevB.108.L201108,2023arXiv230812750K,fabian1,fabian2,fabian3,zhan2024cooperation,PhysRevMaterials.8.074802,wang2024pressure,Xue_2024,Zhang2024,PhysRevB.108.165141,jiang2024high,PhysRevB.108.214522,oh2025highspinlowspin,Geisler2024,geisler2025,khaliullin2025,oh2025dopingspinonemottinsulator,PhysRevB.111.L241102,PhysRevB.108.L180510,PhysRevB.110.094509,zhu2025quantumphasetransitiondriven,PhysRevB.110.L060510,ncbf-9b8m}.

Recent APRES measurements~\cite{Li_2025,yue2025correlatedelectronicstructuresunconventional,wang2025electronicstructurecompressivelystrained} have shed some light on this issue and revealed that the low-energy fermions in La$_3$Ni$_2$O$_7$ have dominant contributions from the $e_g$ subspace of $3d$-orbitals: $d_{x^2-y^2}$ and $d_{z^2}$.
In a recent theoretical study~\cite{wang2025originspinstripesbilayer}, we have  shown that  $\bQ=(\pi/2,\pi/2)$ spin stripe order can stem from a sizable Hund's coupling $J_H$ between the itinerant $d_{x^2-y^2}$ electrons and the more localized spins in the $d_{z^2}$ orbitals.  Furthermore, given the proximity of superconductivity to magnetism, it is likely that the scale of Hund's coupling relative to that of in-plane superexchange will significantly influence the pairing symmetry. Indeed, in Ref.  ~\cite{wang2025originspinstripesbilayer}, we reported that interlayer singlet Cooper pairing was the dominant pairing tendency, which was stabilized by $J_H$, and which is equivalent to an $s$-wave paired state.  
In the case of La$_3$Ni$_2$O$_7$, recent experiments reported signatures of $s$-wave~\cite{fan2025superconductinggapstructurebosonic,guo2025revealingsuperconductinggap} and $d$-wave~\cite{cao2025directobservationdwave} pairing, where a competition between the two is generically expected from the fermiology of the bilayer system~\cite{PhysRevB.84.180513}. From a theoretical perspective, proposals supporting $s$-wave pairing include studies using FRG~\cite{PhysRevB.108.L140505,PhysRevLett.134.076001}, DMRG~\cite{PhysRevLett.132.036502,wang2025originspinstripesbilayer}, mean-field theory~\cite{PhysRevLett.132.146002,PhysRevB.109.165154,PhysRevLett.133.096002,PhysRevB.111.035108,Luo2024} and the fluctuation-exchange approximation~\cite{PhysRevLett.132.106002}, while studies using $t$-$J$ models found $d$-wave pairing similar to that in cuprates~\cite{PhysRevB.108.214522,Jiang_2024,PhysRevLett.132.126503}. 

In this work, we investigate both superconductivity and magnetism in La$_3$Ni$_2$O$_7$ resulting from repulsive interactions including intra-orbital $U$, inter-orbital $U'$, Hund's coupling $J_H$ and pair hopping $J_P\approx J_H$, with the goal of understanding how these microscopic interactions affect the pairing symmetry and the wave vector associated with magnetism.  The effective pairing interaction $V_\text{eff}$ is obtained by {dressing these} bare interactions with particle-hole excitations from spin and orbital fluctuations within the random phase approximation (RPA)~\cite{PhysRevLett.17.433,SCALAPINO1995329,PhysRevB.99.224515,RevModPhys.84.1383}.
We identify three characteristic $s$-wave pairing states that occur upon varying the microscopic interactions: a conventional fully gapped $s$-wave, a fully gapped $s_{\pm}$-wave, and a nodal $s$-wave that interpolates between them. Each requires a sizeable $J_H$ defined relative to the scale of in-plane single electron tunneling.
With weaker $J_H$, these $s$-wave states compete with two $d$-wave states of $x^2-y^2$ ($B_{1g}$) and $xy$ ($B_{2g}$) symmetry, with the $B_{1g}$ occupying a larger region of the phase diagram. Through Hund’s coupling, the interlayer hopping of the $d_{z^2}$ orbitals induces an effective interlayer coupling in the $d_{x^2-y^2}$ sector.
The $d$-wave pairing is stabilized only for small interlayer interaction and hence vanishingly small $J_H$.
Based on these results, we speculate that $s_\pm$-wave pairing is the more likely superconducting ground state in La$_3$Ni$_2$O$_7$.

We then present results for magnetism within the RPA.  We show that even at the non-interacting level, the bare spin and charge susceptibilities are peaked at {$\bm Q = (\pi/2, \pi/2)$}, and that this peak is further strengthened by sizeable Hund's coupling in the RPA. In the regime where Hund's coupling is weak, we find the dominant magnetism is a $( \pi, \pi)$ antiferromagnet whose origin stems from a  band with $d_{z^2}$ character possessing a high density of states that occurs just below the Fermi level.

While these results are obtained in the limit where both orbital degrees of freedom are itinerant, they share some qualitative similarities with earlier DMRG studies~\cite{wang2025originspinstripesbilayer}, which incorporated stronger correlation effects and Mott physics.  Both studies have found $(\pi/2, \pi/2)$ spin stripe order and pairing between electrons across the bilayer, and both resulted from a sizeable Hund's rule coupling.  Despite the drastic differences in assumptions that underlie each of these techniques, such qualitative similarity points towards a robustness of these results that are likely inform experimental observations.

In the following, we present details of our model and method in Sec.~\ref{sec:model}. In Sec.~\ref{sec:results}, we discuss the resulting phase diagram and the corresponding gap symmetries as a function of the bare interactions, as well as the normal state magnetic ordering tendencies in the RPA total spin susceptibility. 
We point out key differences to previous weak-coupling studies, namely the importance of Hund's coupling and the non-degenerate nature of the $e_g$ orbitals.

\begin{figure}
    \includegraphics[width=0.9\linewidth]{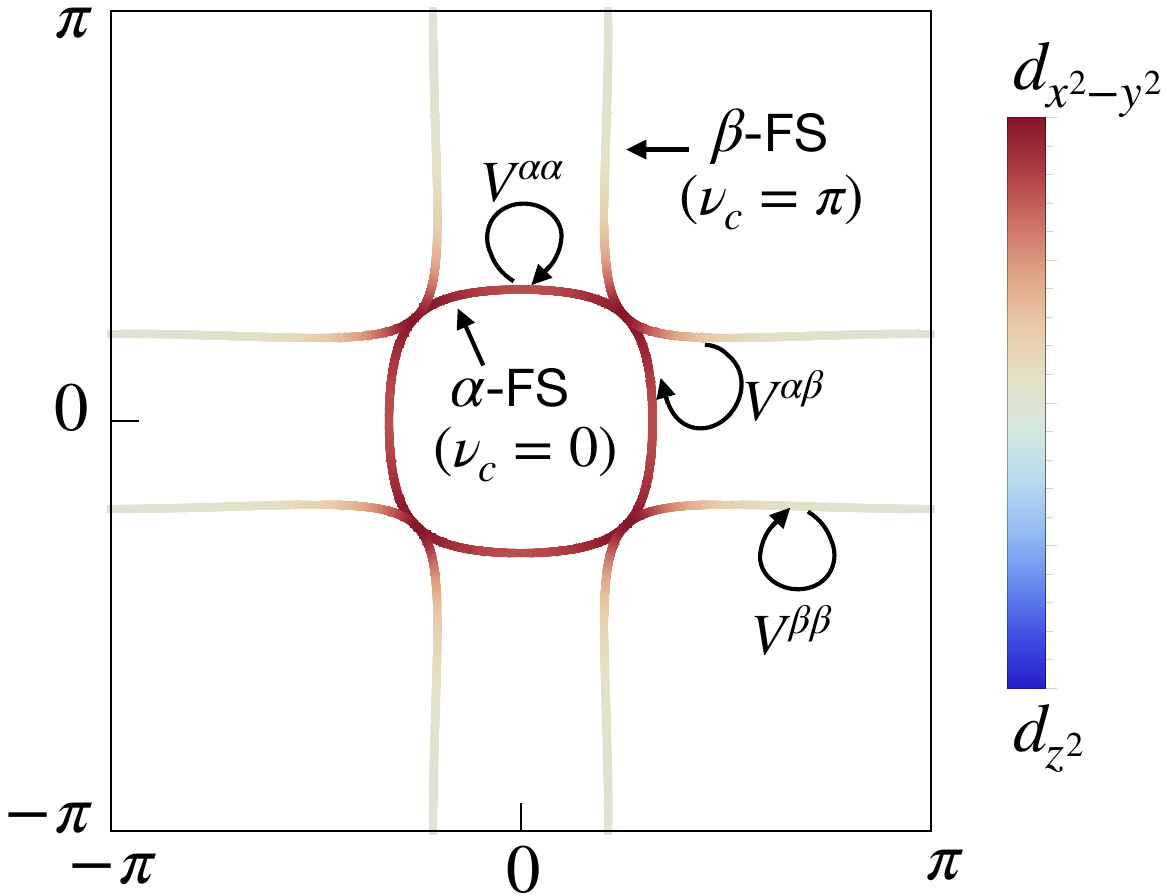}\caption{Fermi surface of bilayer nickelate in the $k_x$-$k_y$ plane from Eq.~\eqref{eq:H00}, with color encoding the orbital weight. Here the central pocket is the $\alpha$-FS (bonding, $\ba = 0$), containing mainly contributions from the $d_{x^2-y^2}$ orbitals. The larger pocket surrounding $(\pi,\pi)$ is the $\beta$-FS (anti-bonding, $\ba = \pi$), and has equal contributions from both orbitals near the BZ boundary. The effective intra-pocket interaction $V^{\alpha\alpha}$ and $V^{\beta\beta}$, and the inter-pocket $V^{\alpha\beta}$ are pictorially shown here, see Fig.~\ref{fig:Valphabeta} for details. }\label{fig:FermiSurface}
\end{figure}

\section{Model and method}  % (fold)
\label{sec:model}
Recent ARPES measurements~\cite{Li_2025,yue2025correlatedelectronicstructuresunconventional,wang2025electronicstructurecompressivelystrained} have indicated that the low energy fermions have dominant contributions from the $e_g$ subspace of the Ni $3d$-orbitals: $d_{x^2-y^2}$ and $d_{z^2}$.
We model the kinetic part of the bilayer nickelates La$_3$Ni$_2$O$_7$ using a tight-binding fit to the ARPES data from Ref. ~\cite{wang2025electronicstructurecompressivelystrained}.
We denote $\psi_{Z}^\dagger,\psi_Z$ for $d_{z^2}$ orbitals and $\psi_X^\dagger, \psi_X$ for $d_{x^2-y^2}$ orbitals.
This results in a spin-degenerate non-interacting Hamiltonian $H_0=\sum_{\bk,\sigma}\psi^\dagger_{\sigma,\bk}\mathcal{H}_0(\bk)\psi_{\sigma,\bk}$ with $\psi^\dagger_{\sigma,\bk}=(\psi^\dagger_{X,\sigma,\bk},\psi^\dagger_{Z,\sigma,\bk})$,
\begin{equation}
     \mathcal{H}_0(\bk)=\begin{pmatrix}
        \xi_{X}(\bk) & T(\bk)\\
        T^\dagger(\bk) & \xi_{Z}(\bk)
    \end{pmatrix}\label{eq:H00}
\end{equation}
and matrix elements
\begin{align}
\label{eq:MatrixElems}
        \xi_X(\bk)&=-2t_{1x}\, (\cos k_x + \cos k_y)-4t_{2x} \cos k_x \cos k_y-\mu_d,\nonumber \\
        \xi_Z(\bk)&=-t_{z}^\bot \cos \ba -2t_{1z} \, (\cos k_x + \cos k_y) \nonumber \\
        &~~~~-4t_{2z} \cos k_x \cos k_y-\mu_f,\nonumber \\
        T(\bk)&= -2t_{xz} \, (\cos k_x-\cos k_y) \nonumber \\
        &~~~~-2t^\bot_{xz}  \cos \ba \,(\cos k_x-\cos k_y).
\end{align}
The values of the parameters are taken from~\cite{wang2025electronicstructurecompressivelystrained} and also given in Appendix~\ref{app:RPA}. {We consider a single bilayer system with bonding and anti-bonding bands within the bilayer corresponding to $\ba=0$ and $\ba=\pi$ respectively. These are the  even ($e^{i \ba} = +1$) and odd components ($e^{i \ba} = -1$) under layer exchange in the $c$-axis bilayer splitting. To incorporate the spatial degrees of freedom, we define $\bk \equiv (k_x,k_y,\ba)$.}
The Fermi surface (FS) obtained from diagonalizing $\mathcal{H}_0(\bk)$ is presented in Fig.~\ref{fig:FermiSurface}. The bonding band ($\ba = 0$) is called $\alpha$-FS and the anti-bonding band ($\ba = \pi$) $\beta$-FS. Note that in the model we use, the $\gamma$-pocket near $(\pi,\pi)$ does not cross the Fermi level, consistent with ARPES data~\cite{wang2025electronicstructurecompressivelystrained} measured on thin film samples.

\begin{figure*}
    \includegraphics[width=\linewidth]{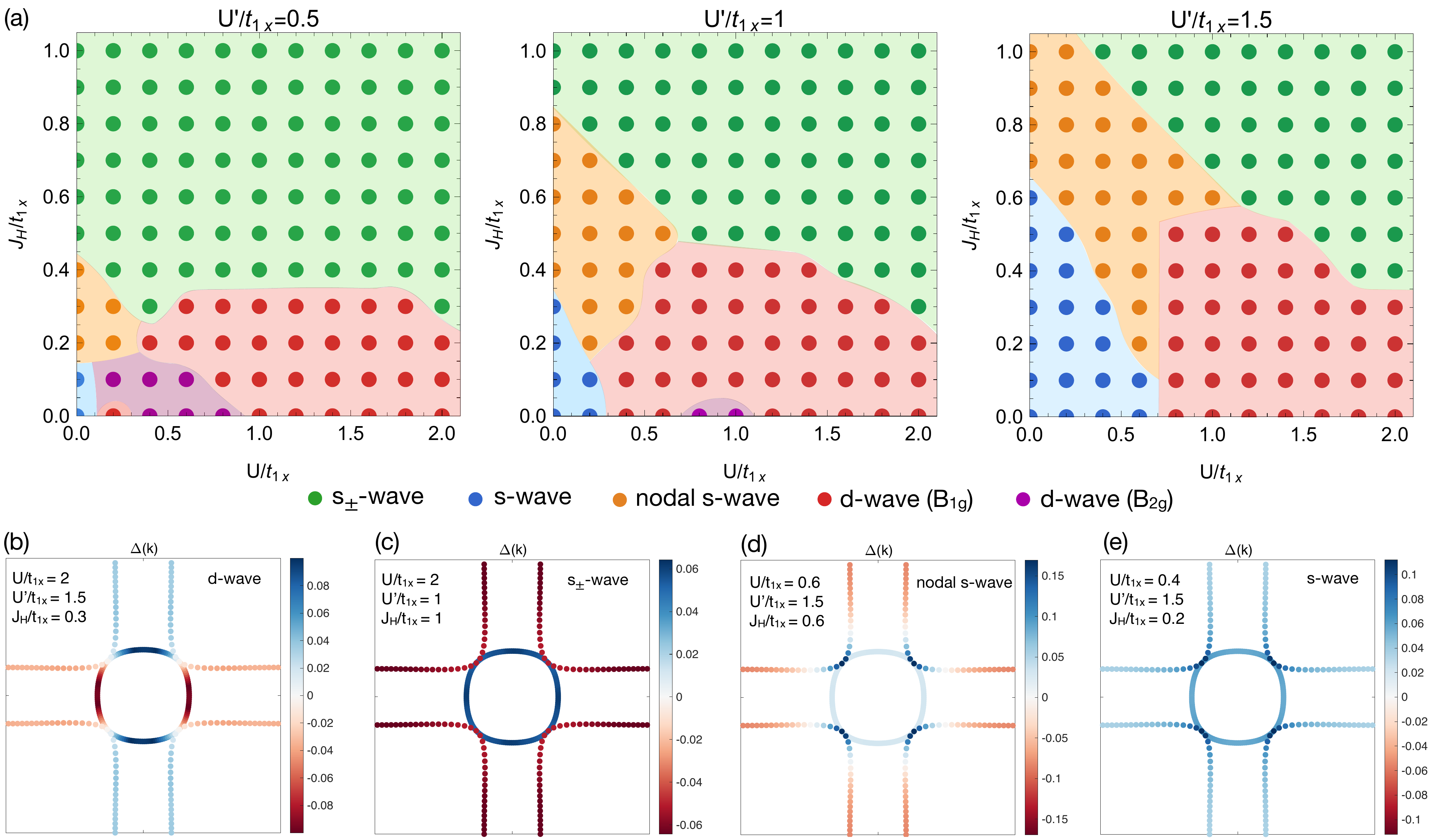}
    \caption{\textbf{(a)} Phase diagram of the leading pairing symmetry in the parameter space $U'$-$U$-$J_H$. \textbf{(b-e)} Band basis projected gap function $\Delta(\bk)$ [see Eq.~\eqref{eq:bandDelta} and Appendix~\ref{app:lin_gap}] obtained at different parameters indicated in the plots. The value on the color bar shows the relative magnitude of the gap function. \textbf{(b)} $d$-wave gap function in the $B_{1g}$ irrep with nodal points along the $\hat x\pm\hat y$ directions. \textbf{(c)} $s_\pm$-wave, \textbf{(d)} nodal $s$-wave and \textbf{(e)} conventional $s$-wave gap functions. }\label{fig:phasediagram}
\end{figure*}

To investigate the magnetic and superconducting instabilities in bilayer nickelates, we must include  interactions via a  Hamiltonian of the form $H=H_0+H_I$.
In a recent DMRG study~\cite{wang2025originspinstripesbilayer}, it was shown that the $\bQ=(\pi/2,\pi/2)$ spin stripe order and a strong interlayer pairing tendency both stem from a sizable Hund's coupling $J_H$ between the itinerant $d_{x^2-y^2}$ electrons and the localized spins in the $d_{z^2}$ orbitals. 
In the itinerant perspective presented here, we allow for charge fluctuations to occur in both  orbitals. Within this approach, one can ask whether a sizeable $J_H$ is similarly needed to obtain the $\bQ=(\pi/2,\pi/2)$ magnetic order and $s$-wave pairing. To investigate this question, we consider the Kanamori form~\cite{kanamori1963} for the interacting part $H_I$ which along with the intra-orbital Hubbard interaction $U$, inter-orbital onsite repulsion $U'$, pair hopping $J_P$, also includes Hund's coupling $J_H$. 
The Hamiltonian is  
\begin{align}\label{eq:interaction}
        H_I&=  U \sum_{\ell,i,a}n_{a,\ell,i,\uparrow}n_{a,\ell,i,\downarrow}+ U'\sum_{a\neq b,\ell,i,\sigma,\sigma'} n_{a,\ell,i,\sigma}n_{b,\ell,i,\sigma'}\nonumber\\
        &~~~-J_H\sum_{\ell,i,\sigma,\sigma'}\psi^\dagger_{X,\ell,i,\sigma}\psi_{X,\ell,i,\sigma'}\psi^\dagger_{Z,\ell,i,\sigma'}\psi_{Z,\ell,i,\sigma}\nonumber\\
        &~~~+ J_P \sum_{\ell,i} \left( \psi^\dagger_{X,\ell,i,\uparrow}\psi^\dagger_{X,\ell,i,\downarrow}\psi_{Z,\ell,i,\downarrow}\psi_{Z,\ell,i,\uparrow}   + h.c. \right).
\end{align}
Here $i$ labels the position of each bilayer {Ni-site}, and $\ell=1,2$ labels the top and bottom layer. $a=X,Z$ is the orbital label, and $\sigma$ is the spin label. The density is defined as $n_{a,\ell,i,\sigma}=\psi^\dagger_{a,\ell,i,\sigma}\psi_{a,\ell,i,\sigma}$.

In the case where the two $e_{g}$ orbitals are degenerate, one may expect from a first order perturbation theory treatment of Coulomb matrix elements that $U'=U-2J_H$~\cite{annurev:/content/journals/10.1146/annurev-conmatphys-020911-125045,khomskii2014transition}. Here, we treat these coupling as unconstrained {\it effective} interactions: we will treat $U'$, $J_H$ and $U$ independently, while keeping $J_P=J_H$. 

We use RPA to explore the possible superconducting pairing instabilities of the effective pairing interaction which is the bare interaction dressed with particle-hole spin and orbital fluctuations and projected into the singlet Cooper channel. We further evaluate the spin susceptibility to shed light on the possible normal state magnetic ordering as a function of the Hund's coupling $J_H$. Within the RPA setup, which is detailed in Appendix~\ref{app:RPA}, we consider the geometric series of diagrams for both transverse and longitudinal spin and orbital fluctuations.

\begin{figure*}
    \centering
    \includegraphics[width=\linewidth]{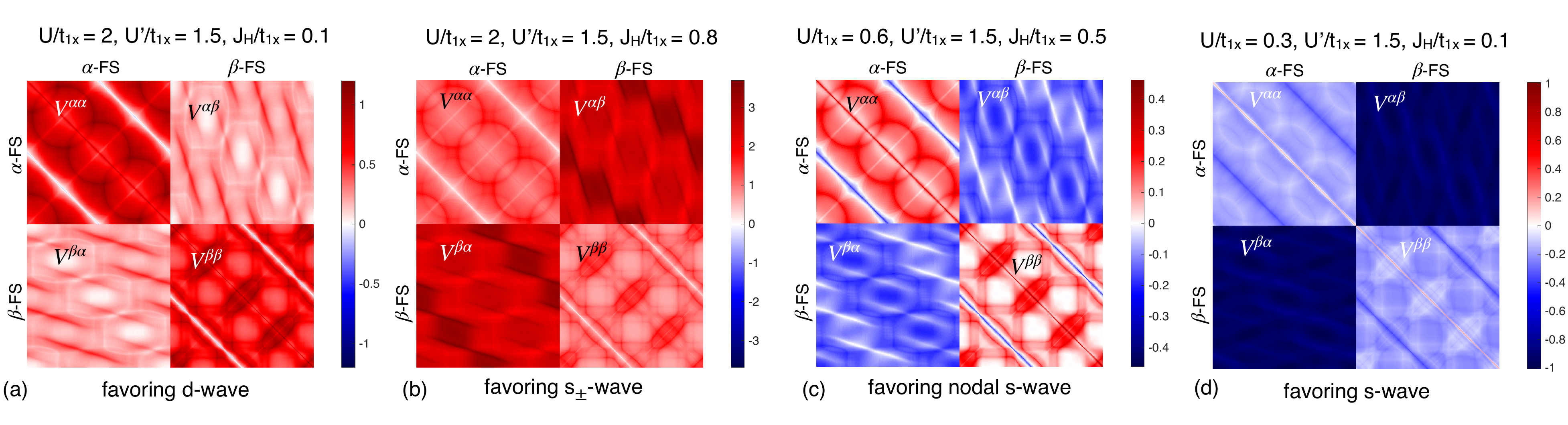}
    \caption{Color maps of the effective interaction $[V_\text{eff}(\bk,\bp)]_{XX,XX}$ between $d_{x^2-y^2}$ orbitals in the $\bk$-$\bp$ plane. Since an equal number of points are sampled on the $\alpha$-FS and on the $\beta$-FS, each plot is divided into four equal-size blocks: the intra-pocket interaction $V^{\alpha\alpha}$ and $V^{\beta\beta}$ and the inter-pocket interaction $V^{\alpha\beta}=[V^{\beta\alpha}]^T$. \textbf{(a)} When all of the interactions are repulsive and the intra-pocket $V^{\alpha\alpha}, V^{\beta\beta}$ dominate, the system favors a $d$-wave. \textbf{(b)} Instead, when the inter-pocket $V^{\alpha\beta}$ dominates in a fully repulsive interaction, the system favors a $s_{\pm}$-wave. \textbf{(c)} When $V^{\alpha\alpha}$ and $V^{\beta\beta}$ are repulsive but $V^{\alpha\beta}$ is attractive, the system favors a nodal $s$-wave. \textbf{(d)} When all these four interactions are attractive, the system favors a conventional $s$-wave.}
    \label{fig:Valphabeta}
\end{figure*}

\section{Results} % (fold)
\label{sec:results}

\subsection{Superconductivity}
We find the leading pairing instability as the solution to the linearized gap equation by the most negative eigenvalue of the symmetrized pairing kernel in the singlet channel, see Appendix~\ref{app:lin_gap} for details. Since the gap function $\Delta_{aa'}$ is a $(2\times 2)$-matrix given in orbital space, we project it to the band basis. Assuming it is the $j$-th band that is on the Fermi surface at each $\bk\in\ $FS, the real-valued band basis projected gap function is given by
\begin{equation}
    \begin{aligned}
        \Delta(\bk) &= \sum_{a,a'} \, u_{a,j}^*(\bk) \, \Delta_{aa'}(\bk) \, u_{a',j}^*(-\bk)
        \end{aligned}\label{eq:bandDelta}
\end{equation}
where $a,a'\in(X, Z)$ and $u_{X,j}^*(\bk)$, $u_{Z,j}^*(\bk)$ are the complex conjugated components of the eigenvector of the $j$-th band of $\mathcal{H}_0(\bk)$.

In the following, we label each solution $\Delta(\bk)$ by the irreducible representation (irrep) of the crystalline symmetry group. 
Within the framework of Landau's theory of phase transitions, any two states $\Delta_1$ and  $\Delta_2$ that belong to the same irrep can mix and coexist. 
In some cases, energetics will favor a state $\Delta_1$ over $\Delta_2$ even though both belong to the same irrep.  This is relevant to the existence of ``accidental nodes" on the gap function.
In such cases, a small $\Delta_2$, although not forbidden, manifests itself at most as a small perturbation to the ground state.  As we discuss below, this is precisely what occurs for various characteristic $s$-wave gap functions as a function of bare interaction strengths.  
Note that any putative phase boundaries between regions with dominant $\Delta_1$, say, and regions with dominant $\Delta_2$  are crossovers, not sharp transitions.  We request the reader to keep this basic framework in mind as we present the results below.

In Fig.~\ref{fig:phasediagram}(a) we show the phase diagrams of the leading pairing symmetries in the $U$-$J_H$ plane for three different {values of} $U'$: $U'/t_{1x}=0.5, 1$ and 1.5. The phase diagram for each $U'$ is plotted as a function of $U/t_{1x} \in [0,2]$ and $J_H/t_{1x} \in [0,1]$.
{We identify five competing pairing channels: fully gapped (nodeless) $s_{\pm}$-wave and $s$-wave, nodal $s$-wave, all of which belong to the same irrep, and $d$-wave states belonging to the $B_{1g}$ and the $B_{2g}$ irreps}. Some representative examples of the band-projected gap function are shown in Fig.~\ref{fig:phasediagram}(b)-(e). 
For each of the  
examples of the band-projected gap function, we plot the effective pairing interaction $V_\text{eff}(\bk,\bk')$ between the $d_{x^2-y^2}$ orbitals projected onto the Fermi surface as color maps in Fig.~\ref{fig:Valphabeta}.

For small Hund's coupling $J_H\lesssim U'/2$, the leading pairing instability occurs in the $B_{1g}$ irrep, corresponding to a $x^2-y^2$-symmetry state, as shown in Fig.~\ref{fig:phasediagram}(b). The gap function has nodes along the diagonal directions ${(\hat x\pm\hat y)}$ and is significantly larger on the $\alpha$-Fermi surface than on the $\beta$-Fermi surface. The effective interaction between $d_{x^2-y^2}$ orbitals is purely repulsive [Fig.~\ref{fig:Valphabeta}(a)] and is dominated by interactions $V^{\alpha\alpha}$, $V^{\beta\beta}$ within each Fermi surface separately. In contrast, inter-pocket scattering $V^{\alpha\beta}$ between the $\alpha$ and $\beta$ sheets is strongly suppressed. This hierarchy follows from the weak interlayer coupling of the $d_{x^2-y^2}$ orbitals at small $J_H$. Since the primary interlayer hopping process involves the $d_{z^2}$ orbitals via the $t_\perp$ hopping, the $d_{x^2-y^2}$ orbitals remain largely confined to individual layers. Consequently, pairing is predominantly intralayer, stabilizing a $B_{1g}$ $d$-wave state.
The nodal structure of this state is controlled by the momentum dependence of the repulsive interaction, dressed by spin fluctuations peaked at $(\pi,\pi)$ as seen in the RPA spin susceptibility (blue) in Fig.~\ref{fig:RPAchi}. While this resembles the spin-fluctuation mechanism underlying $d$-wave pairing in cuprates, here the $(\pi,\pi)$ fluctuations are enabled by the additional $\gamma$ band close to the Fermi level.

Another $d$-wave pairing with $xy$-symmetry in the $B_{2g}$ irrep is leading for vanishingly small $J_H$ and $U\approx U'$. This $d$-wave pairing has nodes only on the $\alpha$-FS along the $\hat x$- and $\hat y$-directions.

At large $J_H$, and approximately for $U+2 \,J_H> 1.6 \,U'$, an $s_\pm$-wave state becomes the leading instability. This state is fully gapped, with opposite signs of the gap function on the $\alpha$- and $\beta$-Fermi surfaces, as shown in Fig.~\ref{fig:phasediagram}(c). For large Hund’s coupling, the interlayer processes active in the $d_{z^2}$ channel generate an effective interlayer coupling in the $d_{x^2-y^2}$ sector. This renormalized interlayer interaction enhances scattering between the $\alpha$ and $\beta$ Fermi surfaces. As reflected in Fig.~\ref{fig:Valphabeta}(b), the effective interaction between the $d_{x^2-y^2}$ orbitals is therefore dominated by inter-pocket repulsion $V^{\alpha\beta}$, while intra-pocket contributions $V^{\alpha\alpha}$, $V^{\beta\beta}$ are suppressed. Such predominantly interband repulsion naturally favors a sign-changing yet fully gapped superconducting state, i.e., an $s_\pm$-pairing symmetry.
The nature of the spin (and orbital) fluctuations that drive the $s_\pm$-wave pairing is also distinct from the $d$-wave case, as with  large $J_H$ the spin fluctuations carry momentum near $(\pi/2,\pi/2)$ instead of $(\pi,\pi)$. This can be seen from the calculated RPA total spin susceptibility in blue in Fig.~\ref{fig:RPAchi}.

When $U+2J_H\lesssim 1.6U'$ and $2J_H>U$, a distinct $s$-wave solution with eight nodal points on the $\beta$ Fermi surface becomes dominant. An example of the corresponding gap function is shown in Fig.~\ref{fig:phasediagram}(d). Starting from the nodeless $s_\pm$-state, nodes first emerge along the diagonal directions and subsequently move toward the Brillouin zone boundary near $(0,\pi)$ and $(\pi,0)$. Upon further reducing $J_H$ and $U$, this nodal $s$-wave continuously crosses over into the conventional nodeless $s$-wave state realized near $J_H=U=0$, as shown in Fig.~\ref{fig:phasediagram}(e). The nodal $s$-wave thus interpolates between the $s_\pm$ and conventional $s$-wave limits. The evolution of the $s$-wave state is reflected in the effective interaction. In the parameter regime favoring the nodal $s$-wave [Fig.~\ref{fig:Valphabeta}(c)], the interaction $V^{\alpha\beta}$ between the $\alpha$- and $\beta$-Fermi surfaces becomes attractive, while the intra-pocket components $V^{\alpha\alpha}$, $V^{\beta\beta}$ remain repulsive. In contrast, in the conventional $s$-wave regime [Fig.~\ref{fig:Valphabeta}(d)], the effective interaction is attractive for both inter- and intra-pocket processes, stabilizing a non-sign-changing $s$-wave gap. We note again that since these $s$-wave states all belong to the same irrep, there will be some admixture of the states; therefore, all nodes within the $s$-wave states are ``accidental", and not protected by symmetry. 

We note that in several previous studies the RPA method has also been applied to inspect the leading pairing symmetries~\cite{PhysRevLett.131.236002,yue2025correlatedelectronicstructuresunconventional,Zhang2024,PhysRevB.108.165141}, with which our result reaches a qualitative agreement. 
However, as we commented above, the Fermi surface we analyze does not include the $\gamma$-pocket near the $(\pi,\pi)$ BZ corner, which is mainly comprised of $d_{z^2}$ orbitals. Thus, the agreement between our findings and previous studies imply the $\gamma$-pocket may be irrelevant to the pairing in bilayer nickelates. 
Finally we note that in our phase diagram Fig.~\ref{fig:phasediagram}(a) the regime for the leading $s_\pm$-wave pairing at $U+2J_H>1.6U'$ can be extended to even stronger interactions, and eventually reach agreement with the strong coupling results~\cite{PhysRevB.110.104517,PhysRevLett.132.036502,wang2025originspinstripesbilayer}.
These results are in qualitative agreement with an earlier DMRG study on the role of Hund's coupling both for superconductivity and magnetism.

\subsection{Magnetism} 
\begin{figure}
    \centering
    \includegraphics[width=\linewidth]{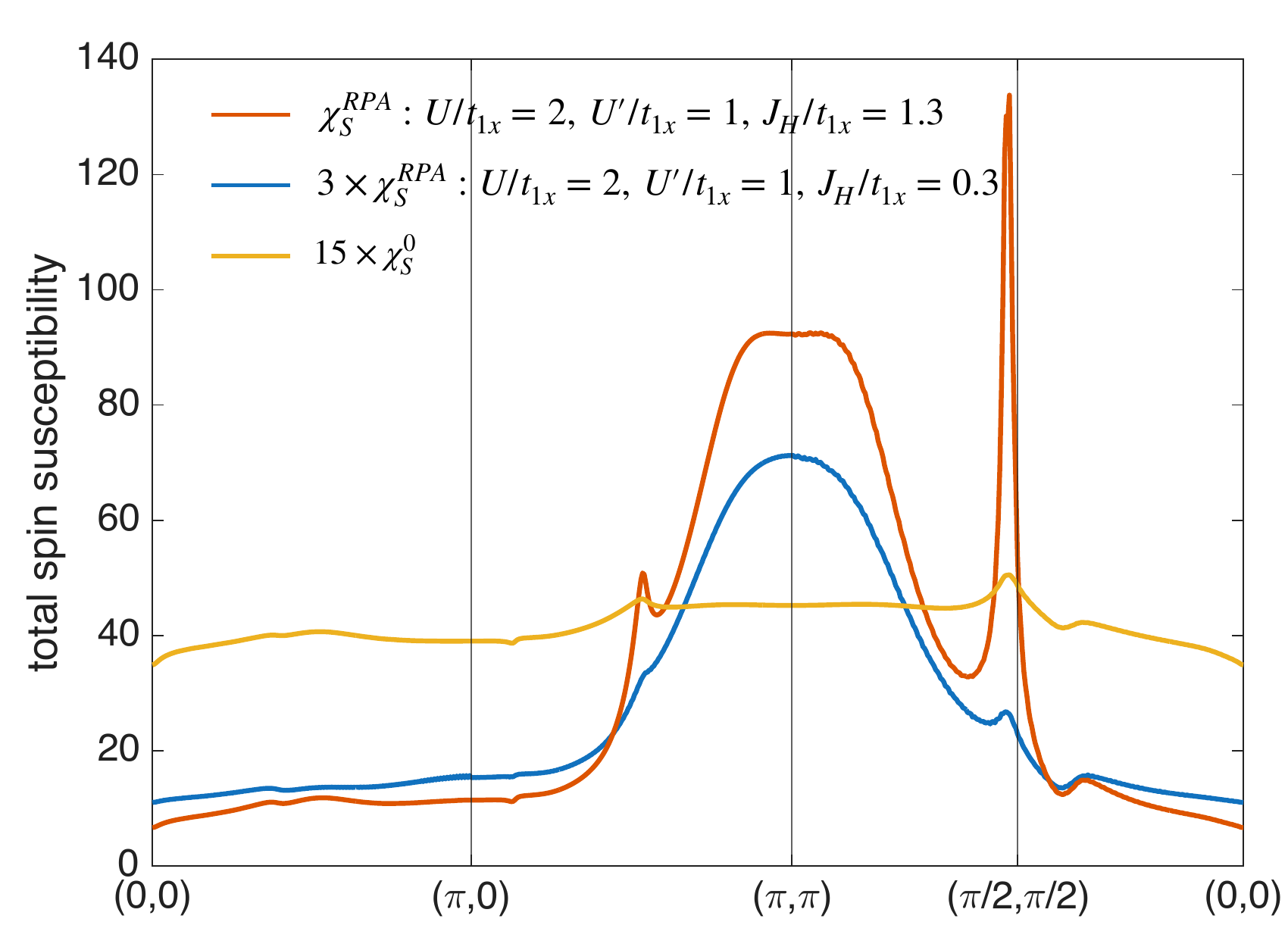}   
    \caption{Comparison of the RPA total spin susceptibility $\chi_S^\text{RPA}(\bq)$ and the bare total spin susceptibility $\chi_S^0(\bq)$, for $\bq$ along the path $\Gamma$--X--M--$\Gamma$ in the BZ. $\chi_S^0(\bq)$ is shown in yellow. For $\chi_S^\text{RPA}(\bq)$, we fix $U$ and $U'$ and show two different values of $J_H$ in blue and orange.}
    \label{fig:RPAchi}
\end{figure}

In order to show how the Hund's coupling $J_H$ qualitatively affects magnetic ordering tendencies in the normal state, we present the results for the total spin susceptibility, calculated within RPA.
The total spin-$S_z$ susceptibility is obtained by 
\begin{equation}
    \chi_S^\text{RPA}(\bq)=\sum_{a,b}\sum_{\alpha\gamma\beta\delta}[\chi^\text{RPA}(\bq)]_{aa\alpha\gamma;bb\beta\delta}\,\sigma^z_{\alpha\gamma}\sigma^z_{\delta\beta},\label{eq:spinRPA}
\end{equation}
where $a,b$ are orbital indices and $\alpha,\beta,\gamma,\delta$ are spin indices.
For comparison, we calculate the bare total spin susceptibility by summing over the orbital indices of the bare susceptibility $\chi^0(\bq)$ (defined in Appendix~\ref{app:RPA}) as
\begin{equation}
    \chi_S^0(\bq)=\sum_{a,b}[\chi^0(\bq)]_{aa,bb}.
\end{equation}

In Fig.~\ref{fig:RPAchi}, we plot the bare and RPA total spin susceptibility along the path $\Gamma$--X--M--$\Gamma$ for $\ba = \pi$. As depicted in yellow in Fig.~\ref{fig:RPAchi}, the bare $\chi_S^0(\bq)$ clearly peaks close to $\bQ=(\pi/2,\pi/2)$, which we trace back to nesting effects between the $\alpha$ and $\beta$ Fermi surface pockets along the diagonal direction in the BZ, see Appendix~\ref{app:susceptibility}.

With the blue and orange curves in Fig.~\ref{fig:RPAchi}, we further compare two different values of $J_H$ of the RPA total spin susceptibility $\chi_S^\text{RPA}(\bq)$ while keeping all other parameters fixed. For a small $J_H$, $\chi_S^\text{RPA}(\bq)$ peaks at momentum $(\pi,\pi)$. The origin of this broad peak may be traced back to off-shell processes involving the $\gamma$-band. Since the $\gamma$-band is only slightly below the Fermi surface, scattering from and into these states, though algebraically suppressed, is still possible provided that interactions are strong enough. We find that among all the included RPA diagrams, the `transverse' fluctuations which is comprised of ladders of $V_1$ do not change the maximum position of $\chi_S^0(\bq)$. It is the `longitudinal' contributions which contain $V_2$ that tend to shift the maximum position to $(\pi,\pi)$~\footnote{See Ref.~\cite{SCALAPINO1995329} for the definition of transverse and longitudinal spin fluctuations.}. At large enough $J_H$, $\chi_S^\text{RPA}(\bq)$ starts to peak near $(\pi/2,\pi/2)$. This is qualitatively consistent with the strong coupling theory of the $(\pi/2,\pi/2)$ order in bilayer nickelates in Ref.~\cite{wang2025originspinstripesbilayer}, where it is shown that a critical value of $J_H$ is needed for the onset of this spin order (see Appendix~\ref{app:susceptibility} for details).

\section{Discussion}  % (fold)
\label{sec:discussion}

Based on the results above, our key inference is that Hund's coupling influences both superconductivity and magnetism of bilayer nickelates in an essential way.  The Hund's coupling favors $s$-wave states, and stabilizes magnetism with wave-vector $(\pi/2, \pi/2)$.  
By locking the $d_{z^2}$ and $d_{x^2-y^2}$ orbital fluctuations together, the Hund's coupling favors a high spin state, which in turn influences the magnetic ordering tendencies.  Furthermore, when $J_H$ couples the $d_{z^2}$ and $d_{x^2-y^2}$ orbital fluctuations, it {\it induces} an interlayer interaction in the $d_{x^2-y^2}$ sector from the {\it intrinsic}  
interlayer coupling of the $d_{z^2}$ orbitals.  In this way, Hund's rule, as well as the bilayer geometry crucially affect pairing. 

We conclude with some broader implications of our findings.  In previous work~\cite{oh2025highspinlowspin}, we reported on high- to low- spin crossovers in putative $d^8$ insulating parent compounds of bilayer nickelates.  Such crossovers occur as the ratio of Hund's coupling to interlayer superexchange between $d_{z^2}$ spins is varied. A large $J_H$ favors a spin-1 ground state (high spin) whereas a large interlayer superexchange favors a low-spin ({\it e.g.} spin 1/2) ground state.  We argue that such high- to low- spin crossovers are also pertinent to the metallic $d^{7.5}$ configuration.  

In the high-spin metallic state, magnetism is enhanced by a strong Hund's coupling, and interlayer superexchange is insufficient to forestall magnetism.  With growing interlayer superexchange, superconducting correlations with Cooper pair formation across the bilayer start to grow.  With further growth of interlayer superexchange, the Hund's coupling is not large enough to lock the $d_{z^2}$ and $d_{x^2-y^2}$ orbital fluctuations and the $d_{z^2}$ orbital forms a quantum disordered ground state comprised of interlayer singlets.  The resulting low spin state can be thought of as weakly coupled layers each with a quarter filled $d_{x^2-y^2}$ band where the $d_{x^2-y^2}$ and $d_{z^2}$ orbitals are nearly decoupled. In this regime, the in-plane pairing interaction dominates for $d_{x^2-y^2}$ orbitals, yielding a $d$-wave superconducting pairing symmetry with intralayer Cooper pair formation.

In Ref.~\cite{wang2025originspinstripesbilayer}, we provided evidence for this scenario by studying a model consisting of ``local moments" in a half-filled $d_{z^2}$ orbital coupled to quarter filled $d_{x^2-y^2}$ conduction electrons via a ferromagnetic ``Kondo" coupling of strength $J_H$.  In that work, we demonstrated that $(\pi/2, \pi/2)$ magnetism forms as a high-spin state stabilized by $J_H$.  We also showed that interlayer Cooper pair formation occurs with increasing interlayer superexchange among $d_{z^2}$ local moments.  Our present RPA theory involves quite different underlying assumptions: both $e_g$ orbitals are partially filled with sizeable charge and spin fluctuations.  Nevertheless, the conclusions of the RPA theory are qualitatively similar to those of Ref.~\cite{wang2025originspinstripesbilayer}. Large $J_H$ favors $(\pi/2, \pi/2)$ magnetism.  Superconductivity from repulsive interactions also favors gaps that change signs between the bonding and anti-bonding Fermi sheets.  These solutions, when viewed in real space consist of interlayer Cooper pair formation from $d_{x^2-y^2}$ electrons. As a result, $J_H$ both stabilizes $(\pi/2,\pi/2)$ magnetism and drives the transition from $d$-wave to $s_\pm$-wave superconductivity.
    
In the real material, neither the local nor itinerant approach adequately captures the behavior of the microscopic degrees of freedom.  Nevertheless, both frameworks make qualitatively similar predictions for magnetism and superconductivity. We therefore conclude that in this system,  high temperature superconductivity likely develops along the crossover from high spin to low spin ground states.

\begin{acknowledgements}
We thank Steven Kivelson, Bai Yang Wang, Jiarui Li and Hanbit Oh for useful discussions. 
Y.-M.W. acknowledges support from
the Gordon and Betty Moore Foundation’s EPiQS Initiative through GBMF8686. 
S.V.S. acknowledges the hospitality and support of the LITP Stanford.
Y.-M.W., T.H. and S.R. are supported by the US Department
of Energy, Office of Basic Energy Sciences, Division of
Materials Sciences and Engineering, under Contract No.
DE-AC02-76SF00515.
S.V.S. was supported by the Hanns-Seidel Foundation and the Three Physicists Philanthropic Trust.
T.H. was supported by the Deutsche Forschungsgemeinschaft (DFG, German Research Foundation) under Project No. 537357978.
\end{acknowledgements}

\appendix

\section{Multiband RPA setup} % (fold)
\label{app:RPA}

We set up the kinetic model of the bilayer nickelates as the single-particle Hamiltonian defined in Eq.~\eqref{eq:MatrixElems} of the main text. It is given by
$H_0=\sum_{\bk,\sigma}\psi^\dagger_{\sigma,\bk}\mathcal{H}_0(\bk)\psi_{\sigma,\bk}$ with $\psi^\dagger_{\sigma,\bk}=(\psi^\dagger_{X,\sigma,\bk},\psi^\dagger_{Z,\sigma,\bk})$,
\begin{equation}
     \mathcal{H}_0(\bk)=\begin{pmatrix}
        \xi_{X}(\bk) & T(\bk)\\
        T^\dagger(\bk) & \xi_{Z}(\bk)
    \end{pmatrix}
\end{equation}
and matrix elements
\begin{align}
        \xi_X(\bk)&=-2t_{1x}\, (\cos k_x + \cos k_y)-4t_{2x} \cos k_x \cos k_y-\mu_d,\nonumber \\
        \xi_Z(\bk)&=-t_{z}^\bot \cos \ba -2t_{1z} \, (\cos k_x + \cos k_y) \nonumber \\
        &~~~~-4t_{2z} \cos k_x \cos k_y-\mu_f,\nonumber \\
        T(\bk)&= -2t_{xz} \, (\cos k_x-\cos k_y) \nonumber \\
        &~~~~-2t^\bot_{xz}  \cos \ba \,(\cos k_x-\cos k_y).
\end{align}
Here, $\ba = 0,\pi$ corresponds to the bonding and antibonding band within the bilayer.
\begin{table}[h]
\centering
\small
\setlength{\tabcolsep}{6pt}
\renewcommand{\arraystretch}{1.2}
\begin{tabular}{|c|c|}
\hline
\textbf{Parameter} & \textbf{Value (eV)} \\
\hline
$t_{1x}$         & $0.345$  \\
$t_{2x}$         & $-0.057$ \\
$\mu_d$          & $-0.851$ \\
$t_{1z}$         & $0.096$  \\
$t_{2z}$         & $0.051$  \\
$t_z^{\bot}$     & $0.42$   \\
$\mu_f$          & $-0.192$ \\
$t_{xz}$         & $0.215$  \\
$t_{xz}^{\bot}$  & $-0.01$  \\
\hline
\end{tabular}
\caption{Model parameters for the tight-binding Hamiltonian in units of eV~\cite{wang2025electronicstructurecompressivelystrained}.}
\label{tab:params}
\end{table}
\begin{figure}[t]
    \includegraphics[width=0.9\linewidth]{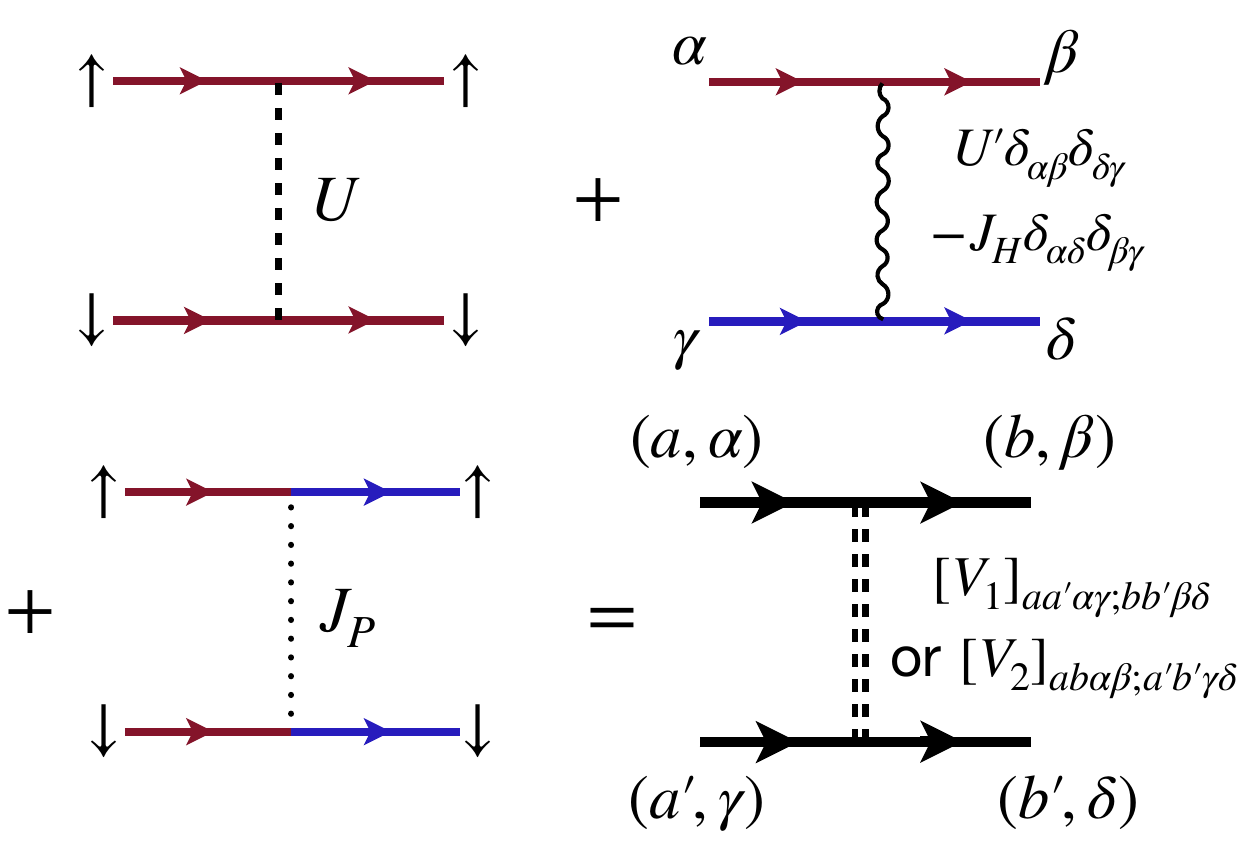}\caption{Diagrammatic representation of the microscopic interactions considered in Eq.~\eqref{eq:interaction_app}. We use red and blue coloring to denote the $d_{x^2-y^2}$ and $d_{z^2}$ orbitals respectively, while the spin indices are explicitly labeled. Depending on how the orbital and spin indices will be contracted, the total interaction can be equivalently written as $V_1$ or $V_2$. Here, the black thick line denotes the matrix propagator including both orbitals.}\label{fig:interactionDiagram}
\end{figure}
The tight-binding parameters we use for the model are stated in Tab.~\ref{tab:params} and are consistent with recent ARPES results~\cite{wang2025electronicstructurecompressivelystrained}. For the interacting part $H_I$ of the Hamiltonian, we use the Kanamori form as stated in Eq.~\eqref{eq:interaction} of the main text,
\begin{align}\label{eq:interaction_app}
    H_I&=  U \sum_{\ell,i,a}n_{a,\ell,i,\uparrow}n_{a,\ell,i,\downarrow}+ U'\sum_{a\neq b,\ell,i,\sigma,\sigma'} n_{a,\ell,i,\sigma}n_{b,\ell,i,\sigma'}\nonumber\\
    &~~~-J_H\sum_{\ell,i,\sigma,\sigma'}\psi^\dagger_{X,\ell,i,\sigma}\psi_{X,\ell,i,\sigma'}\psi^\dagger_{Z,\ell,i,\sigma'}\psi_{Z,\ell,i,\sigma}\nonumber\\
    &~~~+ J_P \sum_{\ell,i} \left( \psi^\dagger_{X,\ell,i,\uparrow}\psi^\dagger_{X,\ell,i,\downarrow}\psi_{Z,\ell,i,\downarrow}\psi_{Z,\ell,i,\uparrow}   + h.c. \right).
\end{align}
We fix $J_P = J_H$, but treat all remaining parameters as independent. The bare interactions of $H_I$ are diagrammatically shown in Fig.~\ref{fig:interactionDiagram}. Note that both $U'$ and the Hund's coupling $J_H$ involve an interaction between the two orbitals, but their action in spin space is different.

In order to study the normal-state magnetic ordering as well as superconductive pairing instabilities, we employ the random phase approximation (RPA). In this method, we evaluate the geometric series of diagrams of both transverse and longitudinal spin and orbital fluctuations. Such methods are standard and have shed considerable insight on unconventional pairing in a variety of materials~\cite{RevModPhys.84.1383}.
Following Fig.~\ref{fig:interactionDiagram}, we label $V_1$ and $V_2$ as interactions arising from the transverse and longitudinal fluctuations, respectively. They correspond to different index contractions of the interaction~\eqref{eq:interaction_app} written as either $[V_1]_{aa'\alpha\gamma;bb'\beta\delta}$ or $[V_2]_{ab\alpha\beta;a'b'\gamma\delta}$, where $aba'b'$ are orbital indices and $\alpha\beta\gamma\delta$ are spin indices. In this orbital-spin basis, both $V_1$ and $V_2$ are $(16\times16)$-matrices, given by
\begin{equation}
    \begin{aligned}
        V_1 & = \begin{pmatrix}
            U & 0 & 0 & J_P \\
            0 & 0 & 0 & 0 \\
            0 & 0 & 0 & 0 \\
            J_P & 0 & 0 & U \\
        \end{pmatrix}\otimes \begin{pmatrix}
            0& 0 & 0 & 0 \\
            0 & 1 & 0 & 0 \\
            0 & 0 & 1 & 0 \\
            0 & 0 & 0 & 0 \\
        \end{pmatrix}\\
        &+ \begin{pmatrix}
            0 & 0 & 0 & 0 \\
            0 & 1 & 0 & 0 \\
            0 & 0 & 1 & 0 \\
            0 & 0 & 0 & 0 \\
        \end{pmatrix}\otimes \begin{pmatrix}
            U'-J_H& 0 & 0 & 0 \\
            0 & U' & -J_H & 0 \\
            0 & -J_H & U' & 0 \\
            0 & 0 & 0 & U'-J_H \\
        \end{pmatrix}
    \end{aligned}
\end{equation}
and
\begin{equation}
    \begin{aligned}
         V_2 & = \begin{pmatrix}
            U & 0 & 0 & 0 \\
            0 & J_P & 0 & 0 \\
            0 & 0 & J_P & 0 \\
            0 & 0 & 0 & U \\
        \end{pmatrix}\otimes \begin{pmatrix}
            0& 0 & 0 & 1 \\
            0 & 0 & 0 & 0 \\
            0 & 0 & 0 & 0 \\
            1 & 0 & 0 & 0 \\
        \end{pmatrix}\\
        &+ \begin{pmatrix}
            0 & 0 & 0 & 1 \\
            0 & 0 & 0 & 0 \\
            0 & 0 & 0 & 0 \\
            1 & 0 & 0 & 0 \\
        \end{pmatrix}\otimes \begin{pmatrix}
            U'-J_H& 0 & 0 & U' \\
            0 & 0 & -J_H & 0 \\
            0 & -J_H & 0 & 0 \\
            U' & 0 & 0 & U'-J_H \\
        \end{pmatrix}.
    \end{aligned}
\end{equation}
The building block of the geometric RPA series is the bare static polarization bubble which, in matrix form, is given by
\begin{equation}
      \chi^0(\bq)=-\int_k G_0(k)\otimes G_0(k+\bq)\otimes \text{Id}_4.\label{eq:chiq}
\end{equation}
Here, $G_0(k)=[i\omega_n-\mathcal{H}_0(\bk)]^{-1}$ is the free matrix Green's function, $\text{Id}_4$ the $(4\times 4)$ identity matrix in spin space and $\int_k\equiv \frac{T}{2}\sum_{\omega_n}\sum_{\ba = 0,\pi}\int\frac{dk_xdk_y}{(2\pi)^2}$. We have used the abbreviation $k=(\omega_n,\bk)$ and $\omega_n=(2n+1)\pi T$ is the fermionic Matsubara frequency.

\begin{figure}[t]
    \includegraphics[width=\linewidth]{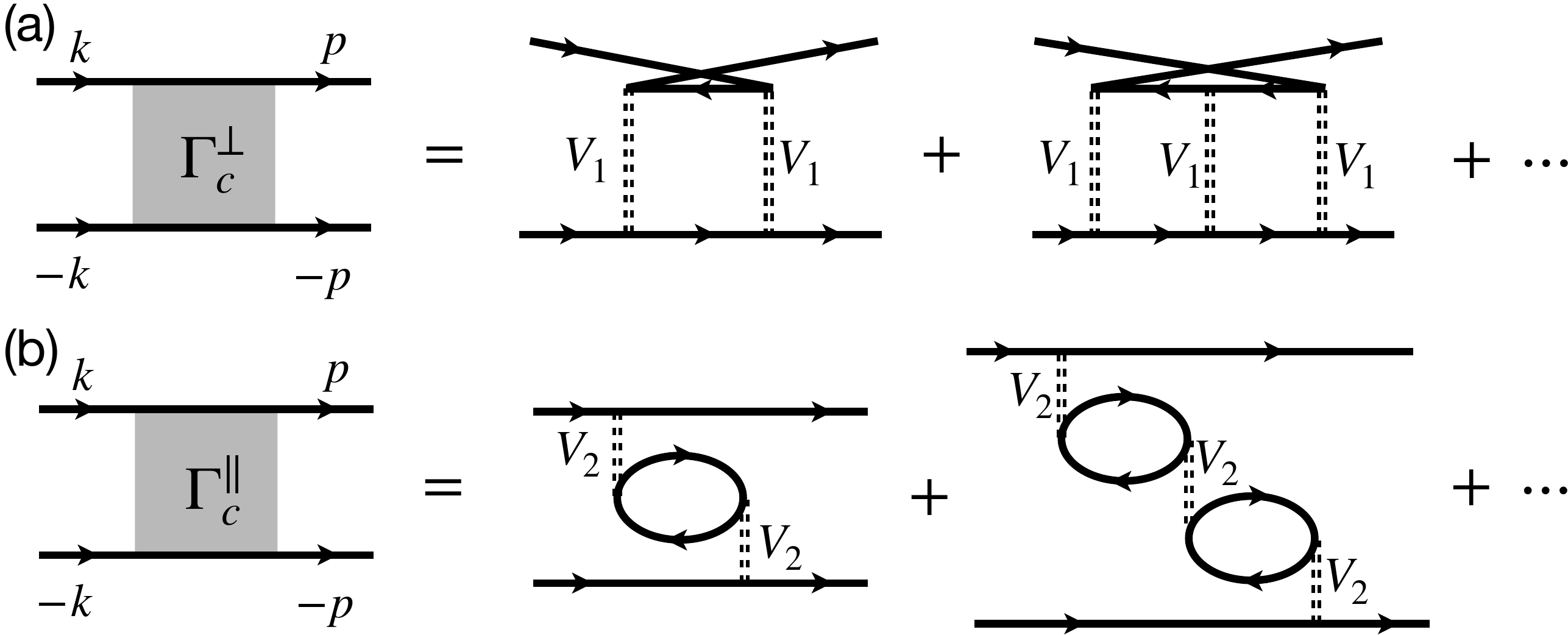}\caption{Effective pairing interaction in the RPA scheme, which contains virtual processes of exchanging  \textbf{(a)} transverse and \textbf{(b)} longitudinal spin-orbital fluctuations. 
    The double dashed line is the tree-level interaction shown in Fig.~\ref{fig:interactionDiagram}.
    }\label{fig:Gammac}
\end{figure}

We obtain the $(16\times 16)$-matrix form of the RPA susceptibility as the sum
\begin{equation}
    \begin{aligned}
        \chi^\text{RPA}(\bq) = \chi_1^\text{RPA}(\bq)  +\chi_2^\text{RPA}(\bq), 
    \end{aligned}
\end{equation}
where the two terms $\chi_1^\text{RPA}(\bq)$ and $\chi_2^\text{RPA}(\bq)$ correspond to the transversal and longitudinal spin-orbital fluctuations respectively, similar to the pairing interaction in Fig.~\ref{fig:Gammac}. They are given by
\begin{equation}
    \begin{aligned}
        %\chi_1^\text{RPA}(\bq)=(1-V_1\chi^0(\bq))^{-1}\chi^0(\bq)\phantom{-}
        \chi_1^\text{RPA}(\bq)= \chi^0(\bq) \,(1-V_1\chi^0(\bq))^{-1}\phantom{-}
    \end{aligned}
\end{equation}
and
\begin{equation}
    \begin{aligned}
        \chi_2^\text{RPA}(\bq)=-\chi^0(\bq) \,(1+V_2\chi^0(\bq))^{-1}.
    \end{aligned}
\end{equation}
Note that the first series is expressed in terms of $V_1$, while the second one is expressed in terms of $V_2$. For the superconducting instability, we then consider the following two geometric series that contribute to the total pairing vertex function $\Gamma_c$ as shown in Fig.~\ref{fig:Gammac},
\begin{equation}
    \begin{aligned}
        %\Gamma_c^\bot(\bk,\bp)&=(1-V_1\chi^0(\bk+\bp))^{-1}V_1\chi^0(\bk+\bp) V_1 \phantom{-}
        \Gamma_c^\bot(\bk,\bp)&=V_1 \, \chi_1^\text{RPA}(\bk+\bp) \, V_1
         \label{eq:gamma_bot}
    \end{aligned}
\end{equation}
and 
\begin{equation}
    \begin{aligned}
        %\Gamma_c^\Vert(\bk,\bp)&=-(1+V_2\chi^0(\bk-\bp))^{-1}V_2\chi^0(\bk-\bp) V_2.
        \Gamma_c^\Vert(\bk,\bp)&=V_2 \, \chi_2^\text{RPA}(\bk-\bp) \, V_2.
        \label{eq:gamma_vert}
    \end{aligned}
\end{equation}

\section{Linearized gap equation} % (fold)
\label{app:lin_gap} 

In the low energy limit, we consider the total spin $S_z=0$ pairing configuration on the Fermi surface, and the effective pairing interaction in this channel is 
\begin{equation}
    \sum_{\bk,\bp}\sum_{aa',bb'}[V_\text{eff}(\bk,\bp)]_{aa',bb'}\psi^\dagger_{a,\uparrow,\bk}\psi^\dagger_{a',\downarrow,-\bk}\psi_{b',\downarrow,-\bp}\psi_{b,\uparrow,\bp}\label{eq:Veff1},
\end{equation}
where 
\begin{equation}
    \begin{aligned}
        [V_\text{eff}(\bk,\bp)]_{aa',bb'}&=[V_1+\Gamma_c^\bot(\bk,\bp)]_{aa'\uparrow\downarrow,bb'\uparrow\downarrow}\\
        &+[\Gamma_c^\Vert(\bk,\bp)]_{ab\uparrow\uparrow,a'b'\downarrow\downarrow}\label{eq:Veff2}
    \end{aligned}
\end{equation}
and both $\bk$ and $\bp$ are on the Fermi surface. 
In our RPA analysis, we explicitly include the bare (tree-level) interaction term in the pairing vertex, as this contribution becomes important whenever a conventional $s$-wave pairing channel competes with fluctuation-mediated pairings.
The most general form of the superconducting order parameter is written as
\begin{equation}
    \hat\Delta(\bk)=\sum_{i=1}^4\Delta_i(\bk)\Gamma^i=\begin{pmatrix}
        \Delta_{XX}(\bk) & \Delta_{XZ}(\bk)\\
        \Delta_{ZX}(\bk) & \Delta_{ZZ}(\bk)    \end{pmatrix}\label{eq:DeltaXZ}
\end{equation}
with 
\begin{equation}
    \Gamma^1=\begin{pmatrix}
        1 & 0\\
        0 & 0
    \end{pmatrix},~ \Gamma^2=\begin{pmatrix}
        0 & 1\\
        0 & 0
    \end{pmatrix},~ \Gamma^3=\begin{pmatrix}
        0 & 0\\
        1 & 0
    \end{pmatrix},~\Gamma^4=\begin{pmatrix}
        0 & 0\\
        0 & 1
    \end{pmatrix}.\nonumber
\end{equation}
Therefore we identify $\Delta_1(\bk)=\Delta_{XX}(\bk)$, $\Delta_2(\bk)=\Delta_{XZ}(\bk)$, $\Delta_3(\bk)=\Delta_{ZX}(\bk)$ and $\Delta_4(\bk)=\Delta_{ZZ}(\bk)$. The subscripts are chosen to encode the orbital information, e.g. $\Delta_{XX}$ is the pairing gap within the $d_{x^2-y^2}$ orbitals.
The linearized gap equation from the effective pairing interaction on the Fermi surface is given by
\begin{equation}
    \Delta_i(\bk)=-\frac{1}{N_\text{FS}}\sum_{\bp\in\text{FS}}{\sum_{j,l=1}^4}[V_\text{eff}(\bk,\bp)]_{i,j}\, P^{j,l}(\bp) \, \Delta_l(\bp),\label{eq:lineargap}
\end{equation}
$N_\text{FS}$ is the number of sampled momentum points on the FS. The particle-particle polarization matrix $P(\bp)$ is given by $P^{j,l}(\bp)=\text{Tr}\left[\frac{T}{N_\bot}\sum_{\bp_\bot,\omega_n}G_0(-p)\Gamma^jG_0(p)\Gamma^l\right]$ where $p=(\omega_n,\bp,\bp_\bot)$ and $\bp_\bot$ denotes $N_\bot$ momentum points along the direction perpendicular to the FS at $\bp$. Therefore, the leading pairing symmetry can be found by diagonalizing the  symmetrized kernel~\cite{PhysRevB.81.224505,PhysRevB.88.064505,PhysRevLett.130.126001}
\begin{equation}
    g_{\bk,\bp}=\frac{1}{N_\text{FS}}P^{\frac{1}{2}}(\bk)V_\text{eff}(\bk,\bp)P^{\frac{1}{2}}(\bp) \label{eq:gkp}
\end{equation}
and identifying the dominant pairing channel as the most negative eigenvalue of~\eqref{eq:gkp}. For that, we use a discretization of 150 points in each of the $\alpha, \beta$ Fermi surface pockets and obtain the static susceptibility $\chi^0(\bq)$ in the $T \rightarrow 0$ limit on a discretized $100 \times 100$ mesh of the BZ. The pair wave function is then given by the eigenvector associated to the dominant pairing strength.  Since the gap function in Eq.~\eqref{eq:DeltaXZ} is defined in terms of the orbital basis, we similarity transform it to the band basis.

Assuming it is the $j$-th band that is on the Fermi surface at each $\bk\in\ $FS, the band basis projected gap function is given by
\begin{equation}
    \begin{aligned}
        \Delta(\bk) &= \sum_{a,a'} \, u_{a,j}^*(\bk) \, \Delta_{aa'}(\bk) \, u_{a',j}^*(-\bk)
        \end{aligned}
\end{equation}
where $a,a'\in(X, Z)$ and $u_{X,j}^*(\bk)$, $u_{Z,j}^*(\bk)$ are the complex conjugated components of the eigenvector of the $j$-th band of $\mathcal{H}_0(\bk)$. Given that the bands of $\mathcal{H}_0(\bk)$ are topologically trivial, we choose the projected gap function $\Delta(\bk)$ to be real.

\section{RPA spin susceptibility}  % (fold)
\label{app:susceptibility}

\begin{figure}
    \centering
    \includegraphics[width=\linewidth]{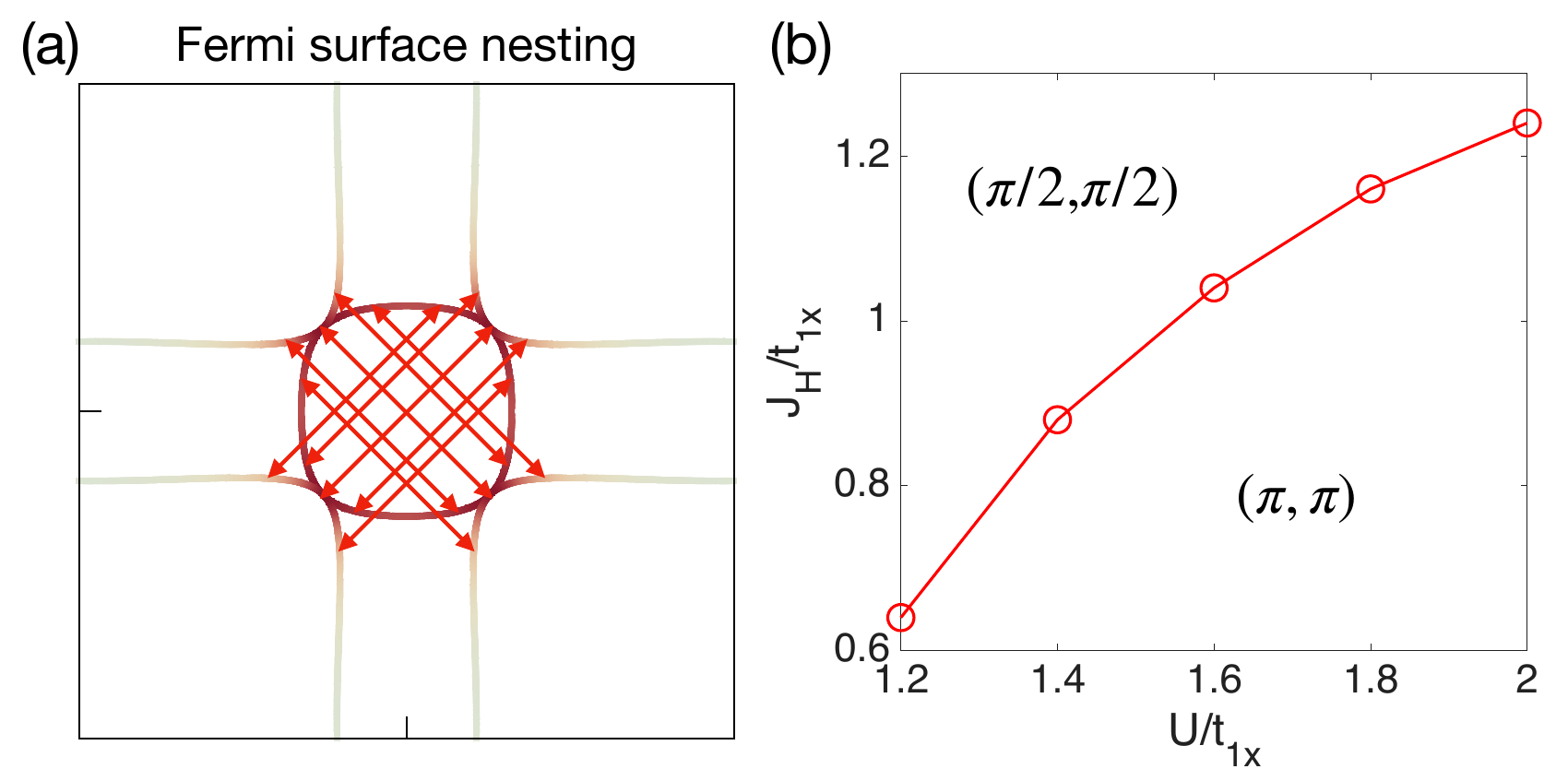}   
    \caption{\textbf{(a)} Fermi surface nesting of momenta close to $(\pi/2,\pi/2)$, as visible in the bare total spin susceptibility $\chi_S^0(\bq)$ in Fig.~\ref{fig:RPAchi} of the main text. \textbf{(b)} Separation between the dominant $(\pi,\pi)$ and $(\pi/2,\pi/2)$ spin fluctuations, determined by the maximum position of the RPA total spin susceptibility $\chi_S^\text{RPA}(\bq)$ in Fig.~\ref{fig:RPAchi} of the main text.}
    \label{fig:Nesting_CriticalJH}
\end{figure}

As discussed in the main text, we consider the bare total spin susceptibility which is obtained by summing over the orbital indices of the bare susceptibility $\chi^0(\bq)$ (defined in Appendix~\ref{app:RPA}) as
\begin{equation}
    \chi_S^0(\bq)=\sum_{a,b}[\chi^0(\bq)]_{aa,bb}.
\end{equation}
The full RPA total spin-$S_z$ susceptibility is then obtained by 
\begin{equation}
    \chi_S^\text{RPA}(\bq)=\sum_{a,b}\sum_{\alpha\gamma\beta\delta}[\chi^\text{RPA}(\bq)]_{aa\alpha\gamma;bb\beta\delta}\,\sigma^z_{\alpha\gamma}\sigma^z_{\delta\beta},\label{eq:spinRPA_app}
\end{equation}
where $a,b$ are orbital indices and $\alpha,\beta,\gamma,\delta$ are spin indices.
In Fig.~\ref{fig:Nesting_CriticalJH}(a) we show nesting effects within the Fermi surface with the red arrows being momentum, measured by the peak positions in $\chi_S^0(\bq)$, as plotted in Fig.~\ref{fig:RPAchi} of the main text. Since we implicitly keep $\ba = \pi$, the momenta have to connect $\alpha$-FS to $\beta$-FS and vice versa. Although not perfect, the nesting effect can still be seen from the figure which mainly occurs between $d_{x^2-y^2}$ orbitals.

In the main text, we discussed the competition between a $(\pi/2,\pi/2)$ and a $(\pi,\pi)$ SDW order. 
%Lastly, the shifting of the peak position from near $(\pi/2,\pi/2)$ to $(\pi,\pi)$ implies a competition between these two dominant spin orders.
In Fig.~\ref{fig:Nesting_CriticalJH}(b) we present the boundary between the two spin orders in the $J_H$-$U$ plane by keeping $U'=t_{1x}$.  We note that our weak coupling theory here suggest that the critical $J_H$ needed to induce the near $(\pi/2,\pi/2)$ order {\it increases} with increasing $U$. This is opposite to the result of the strong coupling limit~\cite{wang2025originspinstripesbilayer}.

\bibliography{nickelate}

@article{kanamori1963,
    author = {Kanamori, Junjiro},
    title = {Electron Correlation and Ferromagnetism of Transition Metals},
    journal = {Progress of Theoretical Physics},
    volume = {30},
    number = {3},
    pages = {275-289},
    year = {1963},
    month = {09},
    issn = {0033-068X},
    doi = {10.1143/PTP.30.275},
    url = {https://doi.org/10.1143/PTP.30.275}
}

@article{PhysRevB.84.180513,
  title = {Pair structure and the pairing interaction in a bilayer Hubbard model for unconventional superconductivity},
  author = {Maier, T. A. and Scalapino, D. J.},
  journal = {Phys. Rev. B},
  volume = {84},
  issue = {18},
  pages = {180513},
  numpages = {4},
  year = {2011},
  month = {Nov},
  publisher = {American Physical Society},
  doi = {10.1103/PhysRevB.84.180513},
  url = {https://link.aps.org/doi/10.1103/PhysRevB.84.180513}
}

@article{PhysRevB.81.224505,
  title = {Superconductivity in the repulsive Hubbard model: An asymptotically exact weak-coupling solution},
  author = {Raghu, S. and Kivelson, S. A. and Scalapino, D. J.},
  journal = {Phys. Rev. B},
  volume = {81},
  issue = {22},
  pages = {224505},
  numpages = {17},
  year = {2010},
  month = {Jun},
  publisher = {American Physical Society},
  doi = {10.1103/PhysRevB.81.224505},
  url = {https://link.aps.org/doi/10.1103/PhysRevB.81.224505}
}

@article{PhysRevB.88.064505,
  title = {Band structure effects on the superconductivity in Hubbard models},
  author = {Cho, Weejee and Thomale, Ronny and Raghu, Srinivas and Kivelson, Steven A.},
  journal = {Phys. Rev. B},
  volume = {88},
  issue = {6},
  pages = {064505},
  numpages = {14},
  year = {2013},
  month = {Aug},
  publisher = {American Physical Society},
  doi = {10.1103/PhysRevB.88.064505},
  url = {https://link.aps.org/doi/10.1103/PhysRevB.88.064505}
}

@article{PhysRevLett.130.126001,
  title = {Pair-Density-Wave and Chiral Superconductivity in Twisted Bilayer Transition Metal Dichalcogenides},
  author = {Wu, Yi-Ming and Wu, Zhengzhi and Yao, Hong},
  journal = {Phys. Rev. Lett.},
  volume = {130},
  issue = {12},
  pages = {126001},
  numpages = {8},
  year = {2023},
  month = {Mar},
  publisher = {American Physical Society},
  doi = {10.1103/PhysRevLett.130.126001},
  url = {https://link.aps.org/doi/10.1103/PhysRevLett.130.126001}
}

@article{RevModPhys.84.1383,
  title = {A common thread: The pairing interaction for unconventional superconductors},
  author = {Scalapino, D. J.},
  journal = {Rev. Mod. Phys.},
  volume = {84},
  issue = {4},
  pages = {1383--1417},
  numpages = {0},
  year = {2012},
  month = {Oct},
  publisher = {American Physical Society},
  doi = {10.1103/RevModPhys.84.1383},
  url = {https://link.aps.org/doi/10.1103/RevModPhys.84.1383}
}

@article{PhysRevB.99.224515,
  title = {{Two pairing domes as Cu$^{2+}$ varies to Cu$^{3+}$}},
  author = {Maier, Thomas and Berlijn, T. and Scalapino, D. J.},
  journal = {Phys. Rev. B},
  volume = {99},
  issue = {22},
  pages = {224515},
  numpages = {4},
  year = {2019},
  month = {Jun},
  publisher = {American Physical Society},
  doi = {10.1103/PhysRevB.99.224515},
  url = {https://link.aps.org/doi/10.1103/PhysRevB.99.224515}
}

@article{SCALAPINO1995329,
title = {{The case for $d_{x^2-y^2}$ pairing in the cuprate superconductors}},
journal = {Physics Reports},
volume = {250},
number = {6},
pages = {329-365},
year = {1995},
issn = {0370-1573},
doi = {https://doi.org/10.1016/0370-1573(94)00086-I},
url = {https://www.sciencedirect.com/science/article/pii/037015739400086I},
author = {D.J. Scalapino}
}

@article{PhysRevLett.17.433,
  title = {Effect of Ferromagnetic Spin Correlations on Superconductivity},
  author = {Berk, N. F. and Schrieffer, J. R.},
  journal = {Phys. Rev. Lett.},
  volume = {17},
  issue = {8},
  pages = {433--435},
  numpages = {0},
  year = {1966},
  month = {Aug},
  publisher = {American Physical Society},
  doi = {10.1103/PhysRevLett.17.433},
  url = {https://link.aps.org/doi/10.1103/PhysRevLett.17.433}
}

@Article{Luo2024,
author={Luo, Zhihui
and Lv, Biao
and Wang, Meng
and W{\'u}, W{\'e}i
and Yao, Dao-Xin},
title={{High-TC superconductivity in La$_3$Ni$_2$O$_7$ based on the bilayer two-orbital t-J model}},
journal={npj Quantum Materials},
year={2024},
month={Aug},
day={13},
volume={9},
number={1},
pages={61},
issn={2397-4648},
doi={10.1038/s41535-024-00668-w},
url={https://doi.org/10.1038/s41535-024-00668-w}
}

@article{Jiang_2024,
doi = {10.1088/0256-307X/41/1/017402},
url = {https://doi.org/10.1088/0256-307X/41/1/017402},
year = {2024},
month = {jan},
publisher = {Chinese Physical Society and IOP Publishing Ltd},
volume = {41},
number = {1},
pages = {017402},
author = {Jiang, Kun and Wang, Ziqiang and Zhang, Fu-Chun},
title = {{High-Temperature Superconductivity in La$_3$Ni$_2$O$_7$}},
journal = {Chinese Physics Letters}
}

@article{PhysRevB.111.035108,
  title = {{$s_{\pm}$-wave superconductivity in the bilayer two-orbital Hubbard model}},
  author = {Zheng, Yao-Yuan and W\'u, W\'ei},
  journal = {Phys. Rev. B},
  volume = {111},
  issue = {3},
  pages = {035108},
  numpages = {7},
  year = {2025},
  month = {Jan},
  publisher = {American Physical Society},
  doi = {10.1103/PhysRevB.111.035108},
  url = {https://link.aps.org/doi/10.1103/PhysRevB.111.035108}
}

@article{PhysRevLett.132.106002,
  title = {{Possible High ${T}_{c}$ Superconductivity in La$_3$Ni$_2$O$_7$ under High Pressure through Manifestation of a Nearly Half-Filled Bilayer Hubbard Model}},
  author = {Sakakibara, Hirofumi and Kitamine, Naoya and Ochi, Masayuki and Kuroki, Kazuhiko},
  journal = {Phys. Rev. Lett.},
  volume = {132},
  issue = {10},
  pages = {106002},
  numpages = {6},
  year = {2024},
  month = {Mar},
  publisher = {American Physical Society},
  doi = {10.1103/PhysRevLett.132.106002},
  url = {https://link.aps.org/doi/10.1103/PhysRevLett.132.106002}
}

@article{PhysRevB.109.165154,
  title = {{Correlation effects and concomitant two-orbital $s_{\pm}$-wave superconductivity in La$_3$Ni$_2$O$_7$ under high pressure}},
  author = {Tian, Yi-Heng and Chen, Yin and Wang, Jia-Ming and He, Rong-Qiang and Lu, Zhong-Yi},
  journal = {Phys. Rev. B},
  volume = {109},
  issue = {16},
  pages = {165154},
  numpages = {6},
  year = {2024},
  month = {Apr},
  publisher = {American Physical Society},
  doi = {10.1103/PhysRevB.109.165154},
  url = {https://link.aps.org/doi/10.1103/PhysRevB.109.165154}
}

@article{PhysRevLett.134.076001,
  title = {{Theory of Pressure Dependence of Superconductivity in Bilayer Nickelate La$_3$Ni$_2$O$_7$}},
  author = {Jiang, Kai-Yue and Cao, Yu-Han and Yang, Qing-Geng and Lu, Hong-Yan and Wang, Qiang-Hua},
  journal = {Phys. Rev. Lett.},
  volume = {134},
  issue = {7},
  pages = {076001},
  numpages = {8},
  year = {2025},
  month = {Feb},
  publisher = {American Physical Society},
  doi = {10.1103/PhysRevLett.134.076001},
  url = {https://link.aps.org/doi/10.1103/PhysRevLett.134.076001}
}

@article{PhysRevLett.59.1613,
  title = {{Two-dimensional antiferromagnetic quantum spin-fluid state in La$_2$CuO$_4$}},
  author = {Shirane, G. and Endoh, Y. and Birgeneau, R. J. and Kastner, M. A. and Hidaka, Y. and Oda, M. and Suzuki, M. and Murakami, T.},
  journal = {Phys. Rev. Lett.},
  volume = {59},
  issue = {14},
  pages = {1613--1616},
  numpages = {0},
  year = {1987},
  month = {Oct},
  publisher = {American Physical Society},
  doi = {10.1103/PhysRevLett.59.1613},
  url = {https://link.aps.org/doi/10.1103/PhysRevLett.59.1613}
}

@article{PhysRevB.37.7443,
  title = {{Static and dynamic spin correlations in pure and doped La$_2$CuO$_4$}},
  author = {Endoh, Y. and Yamada, K. and Birgeneau, R. J. and Gabbe, D. R. and Jenssen, H. P. and Kastner, M. A. and Peters, C. J. and Picone, P. J. and Thurston, T. R. and Tranquada, J. M. and Shirane, G. and Hidaka, Y. and Oda, M. and Enomoto, Y. and Suzuki, M. and Murakami, T.},
  journal = {Phys. Rev. B},
  volume = {37},
  issue = {13},
  pages = {7443--7453},
  numpages = {0},
  year = {1988},
  month = {May},
  publisher = {American Physical Society},
  doi = {10.1103/PhysRevB.37.7443},
  url = {https://link.aps.org/doi/10.1103/PhysRevB.37.7443}
}

@misc{wang2025originspinstripesbilayer,
      title={{Origin of Spin Stripes in Bilayer Nickelate La$_3$Ni$_2$O$_7$}}, 
      author={Hao-Xin Wang and Hanbit Oh and Tobias Helbig and Bai Yang Wang and Jiarui Li and Yijun Yu and Harold Y. Hwang and Hong-Chen Jiang and Yi-Ming Wu and S. Raghu},
      year={2025},
      eprint={2509.25344},
      archivePrefix={arXiv},
      primaryClass={cond-mat.supr-con},
      url={https://arxiv.org/abs/2509.25344}, 
}

@Article{Gupta2025,
author={Gupta, Naman K.
and Gong, Rantong
and Wu, Yi
and Kang, Mingu
and Parzyck, Christopher T.
and Gregory, Benjamin Z.
and Costa, Noah
and Sutarto, Ronny
and Sarker, Suchismita
and Singer, Andrej
and Schlom, Darrell G.
and Shen, Kyle M.
and Hawthorn, David G.},
title={{Anisotropic spin stripe domains in bilayer La$_3$Ni$_2$O$_7$}},
journal={Nature Communications},
year={2025},
month={Jul},
day={16},
volume={16},
number={1},
pages={6560},
issn={2041-1723},
doi={10.1038/s41467-025-61653-w},
url={https://doi.org/10.1038/s41467-025-61653-w}
}

@article{PhysRevB.110.L060510,
  title = {Electronic structure, self-doping, and superconducting instability in the alternating single-layer trilayer stacking nickelates ${\mathrm{La}}_{3}{\mathrm{Ni}}_{2}{\mathrm{O}}_{7}$},
  author = {Zhang, Yang and Lin, Ling-Fang and Moreo, Adriana and Maier, Thomas A. and Dagotto, Elbio},
  journal = {Phys. Rev. B},
  volume = {110},
  issue = {6},
  pages = {L060510},
  numpages = {7},
  year = {2024},
  month = {Aug},
  publisher = {American Physical Society},
  doi = {10.1103/PhysRevB.110.L060510},
  url = {https://link.aps.org/doi/10.1103/PhysRevB.110.L060510}
}

@misc{zhu2025quantumphasetransitiondriven,
      title={Quantum phase transition driven by competing intralayer and interlayer hopping of Ni-$d_{3z^2-r^2}$ orbitals in bilayer nickelates}, 
      author={Xiaoyu Zhu and Wei Qin and Ping Cui and Zhenyu Zhang},
      year={2025},
      eprint={2507.11169},
      archivePrefix={arXiv},
      primaryClass={cond-mat.str-el},
      url={https://arxiv.org/abs/2507.11169}, 
}

@article{PhysRevB.110.094509,
  title = {Interplay of two ${E}_{g}$ orbitals in superconducting ${\mathrm{La}}_{3}{\mathrm{Ni}}_{2}{\mathrm{O}}_{7}$ under pressure},
  author = {Lu, Chen and Pan, Zhiming and Yang, Fan and Wu, Congjun},
  journal = {Phys. Rev. B},
  volume = {110},
  issue = {9},
  pages = {094509},
  numpages = {16},
  year = {2024},
  month = {Sep},
  publisher = {American Physical Society},
  doi = {10.1103/PhysRevB.110.094509},
  url = {https://link.aps.org/doi/10.1103/PhysRevB.110.094509}
}

@article{PhysRevB.108.L180510,
  title = {Electronic structure, dimer physics, orbital-selective behavior, and magnetic tendencies in the bilayer nickelate superconductor ${\mathrm{La}}_{3}{\mathrm{Ni}}_{2}{\mathrm{O}}_{7}$ under pressure},
  author = {Zhang, Yang and Lin, Ling-Fang and Moreo, Adriana and Dagotto, Elbio},
  journal = {Phys. Rev. B},
  volume = {108},
  issue = {18},
  pages = {L180510},
  numpages = {5},
  year = {2023},
  month = {Nov},
  publisher = {American Physical Society},
  doi = {10.1103/PhysRevB.108.L180510},
  url = {https://link.aps.org/doi/10.1103/PhysRevB.108.L180510}
}

@article{PhysRevB.111.L241102,
  title = {Strong pairing and symmetric pseudogap metal in a double Kondo lattice model: From a nickelate superconductor to a tetralayer optical lattice},
  author = {Yang, Hui and Oh, Hanbit and Zhang, Ya-Hui},
  journal = {Phys. Rev. B},
  volume = {111},
  issue = {24},
  pages = {L241102},
  numpages = {6},
  year = {2025},
  month = {Jun},
  publisher = {American Physical Society},
  doi = {10.1103/PhysRevB.111.L241102},
  url = {https://link.aps.org/doi/10.1103/PhysRevB.111.L241102}
}

@misc{oh2025dopingspinonemottinsulator,
      title={Doping a spin-one Mott insulator: possible application to bilayer nickelate}, 
      author={Hanbit Oh and Hui Yang and Ya-Hui Zhang},
      year={2025},
      eprint={2509.02673},
      archivePrefix={arXiv},
      primaryClass={cond-mat.str-el},
      url={https://arxiv.org/abs/2509.02673}, 
}

@misc{khaliullin2025,
      title={Orbital Order and Superconductivity in Bilayer Nickelate Compounds}, 
      author={Giniyat Khaliullin and Jiří Chaloupka},
      year={2025},
      eprint={2506.16360},
      archivePrefix={arXiv},
      primaryClass={cond-mat.str-el},
      url={https://arxiv.org/abs/2506.16360}, 
}

@misc{geisler2025,
      title={{Electronic reconstruction and interface engineering of emergent spin fluctuations in compressively strained La$_3$Ni$_2$O$_7$ on SrLaAlO$_4$(001)}}, 
      author={Benjamin Geisler and James J. Hamlin and Gregory R. Stewart and Richard G. Hennig and P. J. Hirschfeld},
      year={2025},
      eprint={2503.10902},
      archivePrefix={arXiv},
      primaryClass={cond-mat.supr-con},
      url={https://arxiv.org/abs/2503.10902}, 
}

@Article{Geisler2024,
author={Geisler, Benjamin
and Fanfarillo, Laura
and Hamlin, James J.
and Stewart, Gregory R.
and Hennig, Richard G.
and Hirschfeld, P. J.},
title={{Optical properties and electronic correlations in La$_3$Ni$_2$O$_7$ bilayer nickelates under high pressure}},
journal={npj Quantum Materials},
year={2024},
month={Nov},
day={14},
volume={9},
number={1},
pages={89},
issn={2397-4648},
doi={10.1038/s41535-024-00690-y},
url={https://doi.org/10.1038/s41535-024-00690-y}
}

@Article{Liu2025,
author={Liu, Yidi
and Ko, Eun Kyo
and Tarn, Yaoju
and Bhatt, Lopa
and Li, Jiarui
and Thampy, Vivek
and Goodge, Berit H.
and Muller, David A.
and Raghu, Srinivas
and Yu, Yijun
and Hwang, Harold Y.},
title={{Superconductivity and normal-state transport in compressively strained La$_2$PrNi$_2$O$_7$ thin films}},
journal={Nature Materials},
year={2025},
month={Aug},
day={01},
volume={24},
number={8},
pages={1221-1227},
issn={1476-4660},
doi={10.1038/s41563-025-02258-y},
url={https://doi.org/10.1038/s41563-025-02258-y}
}

@Article{Zhou2025,
author={Zhou, Guangdi
and Lv, Wei
and Wang, Heng
and Nie, Zihao
and Chen, Yaqi
and Li, Yueying
and Huang, Haoliang
and Chen, Wei-Qiang
and Sun, Yu-Jie
and Xue, Qi-Kun
and Chen, Zhuoyu},
title={{Ambient-pressure superconductivity onset above 40{\thinspace}K in (La,Pr)$_3$Ni$_2$O$_7$ films}},
journal={Nature},
year={2025},
month={Apr},
day={01},
volume={640},
number={8059},
pages={641-646},
issn={1476-4687},
doi={10.1038/s41586-025-08755-z},
url={https://doi.org/10.1038/s41586-025-08755-z}
}

@misc{oh2025highspinlowspin,
      title={High spin, low spin or gapped spins: magnetism in the bilayer nickelates}, 
      author={Hanbit Oh and Yi-Ming Wu and Julian May-Mann and Yijun Yu and Harold Y. Hwang and Ya-Hui Zhang and S. Raghu},
      year={2025},
      eprint={2506.18973},
      archivePrefix={arXiv},
      primaryClass={cond-mat.str-el},
      url={https://arxiv.org/abs/2506.18973}, 
}

@article{kakoi2024multiband,
  title={{Multiband metallic ground state in multilayered nickelates La$_3$Ni$_2$O$_7$ and La$_4$Ni$_3$O$_{10}$ probed by $^{139}$La-NMR at ambient pressure}},
  author={Kakoi, Masataka and Oi, Takashi and Ohshita, Yujiro and Yashima, Mitsuharu and Kuroki, Kazuhiko and Kato, Takeru and Takahashi, Hidefumi and Ishiwata, Shintaro and Adachi, Yoshinobu and Hatada, Naoyuki and others},
  journal={Journal of the Physical Society of Japan},
  volume={93},
  number={5},
  pages={053702},
  year={2024},
  publisher={The Physical Society of Japan}
}

@misc{fan2025superconductinggapstructurebosonic,
      title={{Superconducting gap structure and bosonic mode in La$_2$PrNi$_2$O$_7$ thin films at ambient pressure}}, 
      author={Shengtai Fan and Mengjun Ou and Marius Scholten and Qing Li and Zhiyuan Shang and Yi Wang and Jiasen Xu and Huan Yang and Ilya M. Eremin and Hai-Hu Wen},
      year={2025},
      eprint={2506.01788},
      archivePrefix={arXiv},
      primaryClass={cond-mat.supr-con},
      url={https://arxiv.org/abs/2506.01788}, 
}

@Article{Khasanov2025,
author={Khasanov, Rustem
and Hicken, Thomas J.
and Gawryluk, Dariusz J.
and Sazgari, Vahid
and Plokhikh, Igor
and Sorel, Loïc Pierre
and Bartkowiak, Marek
and Bötzel, Steffen
and Lechermann, Frank
and Eremin, Ilya M.
and Luetkens, Hubertus
and Guguchia, Zurab},
title={{Pressure-enhanced splitting of density wave transitions in La$_3$Ni$_2$O$_{7-\delta}$}},
journal={Nature Physics},
year={2025},
month={Mar},
day={01},
volume={21},
number={3},
pages={430-436},
issn={1745-2481},
doi={10.1038/s41567-024-02754-z},
url={https://doi.org/10.1038/s41567-024-02754-z}
}

@article{XIE20243221,
title = {{Strong interlayer magnetic exchange coupling in La$_3$Ni$_2$O$_{7-\delta}$ revealed by inelastic neutron scattering}},
journal = {Science Bulletin},
volume = {69},
number = {20},
pages = {3221-3227},
year = {2024},
issn = {2095-9273},
doi = {https://doi.org/10.1016/j.scib.2024.07.030},
url = {https://www.sciencedirect.com/science/article/pii/S2095927324005164},
author = {Tao Xie and Mengwu Huo and Xiaosheng Ni and Feiran Shen and Xing Huang and Hualei Sun and Helen C. Walker and Devashibhai Adroja and Dehong Yu and Bing Shen and Lunhua He and Kun Cao and Meng Wang}
}

@article{ZHAO20251239,
title = {{Pressure-enhanced spin-density-wave transition in double-layer nickelate La$_3$Ni$_2$O$_{7-\delta}$ }},
journal = {Science Bulletin},
volume = {70},
number = {8},
pages = {1239-1245},
year = {2025},
issn = {2095-9273},
doi = {https://doi.org/10.1016/j.scib.2025.02.019},
url = {https://www.sciencedirect.com/science/article/pii/S2095927325001811},
author = {Dan Zhao and Yanbing Zhou and Mengwu Huo and Yu Wang and Linpeng Nie and Ye Yang and Jianjun Ying and Meng Wang and Tao Wu and Xianhui Chen}
}

@Article{Wang2024,
author={Wang, Luhong
and Li, Yan
and Xie, Sheng-Yi
and Liu, Fuyang
and Sun, Hualei
and Huang, Chaoxin
and Gao, Yang
and Nakagawa, Takeshi
and Fu, Boyang
and Dong, Bo
and Cao, Zhenhui
and Yu, Runze
and Kawaguchi, Saori I.
and Kadobayashi, Hirokazu
and Wang, Meng
and Jin, Changqing
and Mao, Ho-kwang
and Liu, Haozhe},
title={{Structure Responsible for the Superconducting State in La$_3$Ni$_2$O$_7$ at High-Pressure and Low-Temperature Conditions}},
journal={Journal of the American Chemical Society},
year={2024},
month={Mar},
day={20},
publisher={American Chemical Society},
volume={146},
number={11},
pages={7506-7514},
issn={0002-7863},
doi={10.1021/jacs.3c13094},
url={https://doi.org/10.1021/jacs.3c13094}
}

@book{khomskii2014transition,
  title={Transition metal compounds},
  author={Khomskii, Daniel I},
  year={2014},
  publisher={Cambridge University Press}
}

@article{PhysRevB.108.174511,
  title = {{Type-II $t\ensuremath{-}J$ model and shared superexchange coupling from Hund's rule in superconducting ${\mathrm{La}}_{3}{\mathrm{Ni}}_{2}{\mathrm{O}}_{7}$}},
  author = {Oh, Hanbit and Zhang, Ya-Hui},
  journal = {Phys. Rev. B},
  volume = {108},
  issue = {17},
  pages = {174511},
  numpages = {8},
  year = {2023},
  month = {Nov},
  publisher = {American Physical Society},
  doi = {10.1103/PhysRevB.108.174511},
  url = {https://link.aps.org/doi/10.1103/PhysRevB.108.174511}
}

@misc{yue2025correlatedelectronicstructuresunconventional,
      title={Correlated electronic structures and unconventional superconductivity in bilayer nickelate heterostructures}, 
      author={Changming Yue and Jian-Jian Miao and Haoliang Huang and Yichen Hua and Peng Li and Yueying Li and Guangdi Zhou and Wei Lv and Qishuo Yang and Hongyi Sun and Yu-Jie Sun and Junhao Lin and Qi-Kun Xue and Zhuoyu Chen and Wei-Qiang Chen},
      year={2025},
      eprint={2501.06875},
      archivePrefix={arXiv},
      primaryClass={cond-mat.str-el},
      url={https://arxiv.org/abs/2501.06875}, 
}

@Article{Ren2025,
author={Ren, Xiaolin
and Sutarto, Ronny
and Wu, Xianxin
and Zhang, Jianfeng
and Huang, Hai
and Xiang, Tao
and Hu, Jiangping
and Comin, Riccardo
and Zhou, Xingjiang
and Zhu, Zhihai},
title={{Resolving the electronic ground state of La$_3$Ni$_2$O$_{7-\delta}$ films}},
journal={Communications Physics},
year={2025},
month={Feb},
day={03},
volume={8},
number={1},
pages={52},
issn={2399-3650},
doi={10.1038/s42005-025-01971-z},
url={https://doi.org/10.1038/s42005-025-01971-z}
}

@misc{bhatt2025resolvingstructuraloriginssuperconductivity,
      title={Resolving Structural Origins for Superconductivity in Strain-Engineered La$_3$Ni$_2$O$_7$ Thin Films}, 
      author={Lopa Bhatt and Abigail Y. Jiang and Eun Kyo Ko and Noah Schnitzer and Grace A. Pan and Dan Ferenc Segedin and Yidi Liu and Yijun Yu and Yi-Feng Zhao and Edgar Abarca Morales and Charles M. Brooks and Antia S. Botana and Harold Y. Hwang and Julia A. Mundy and David A. Muller and Berit H. Goodge},
      year={2025},
      eprint={2501.08204},
      archivePrefix={arXiv},
      primaryClass={cond-mat.supr-con},
      url={https://arxiv.org/abs/2501.08204}, 
}

@article{annurev:/content/journals/10.1146/annurev-conmatphys-020911-125045,
   author = "Georges, Antoine and Medici, Luca de&apos; and Mravlje, Jernej",
   title = "Strong Correlations from Hund’s Coupling", 
   journal= "Annual Review of Condensed Matter Physics",
   year = "2013",
   volume = "4",
   number = "Volume 4, 2013",
   pages = "137-178",
   doi = "https://doi.org/10.1146/annurev-conmatphys-020911-125045",
   url = "https://www.annualreviews.org/content/journals/10.1146/annurev-conmatphys-020911-125045",
   publisher = "Annual Reviews",
   issn = "1947-5462",
   type = "Journal Article",
  }

@Article{Ko2024,
author={Ko, Eun Kyo
and Yu, Yijun
and Liu, Yidi
and Bhatt, Lopa
and Li, Jiarui
and Thampy, Vivek
and Kuo, Cheng-Tai
and Wang, Bai Yang
and Lee, Yonghun
and Lee, Kyuho
and Lee, Jun-Sik
and Goodge, Berit H.
and Muller, David A.
and Hwang, Harold Y.},
title={{Signatures of ambient pressure superconductivity in thin film La$_3$Ni$_2$O$_7$}},
journal={Nature},
year={2024},
month={Dec},
day={19},
issn={1476-4687},
doi={10.1038/s41586-024-08525-3},
url={https://doi.org/10.1038/s41586-024-08525-3}
}

@article{PhysRevLett.131.126001,
  title = {Bilayer Two-Orbital Model of $\mathrm{L}{\mathrm{a}}_{3}\mathrm{N}{\mathrm{i}}_{2}{\mathrm{O}}_{7}$ under Pressure},
  author = {Luo, Zhihui and Hu, Xunwu and Wang, Meng and W\'u, W\'ei and Yao, Dao-Xin},
  journal = {Phys. Rev. Lett.},
  volume = {131},
  issue = {12},
  pages = {126001},
  numpages = {6},
  year = {2023},
  month = {Sep},
  publisher = {American Physical Society},
  doi = {10.1103/PhysRevLett.131.126001},
  url = {https://link.aps.org/doi/10.1103/PhysRevLett.131.126001}
}

@article{Sun2023,
  author = {Hualei Sun and Mengwu Huo and Xunwu Hu and Jingyuan Li and Zengjia Liu and Yifeng Han and Lingyun Tang and Zhongquan Mao and Pengtao Yang and Bosen Wang and Jinguang Cheng and Dao-Xin Yao and Guang-Ming Zhang and Meng Wang},
  title = {Signatures of superconductivity near 80 K in a nickelate under high pressure},
  journal = {Nature},
  volume = {621},
  pages = {493--498},
  year = {2023},
  doi = {10.1038/s41586-023-06408-7},
  url = {https://www.nature.com/articles/s41586-023-06408-7}
}

@article{Zhang2024exp,
  author = {Yanan Zhang and Dajun Su and Yanen Huang and Zhaoyang Shan and Hualei Sun and Mengwu Huo and Kaixin Ye and Jiawen Zhang and Zihan Yang and Yongkang Xu and Yi Su and Rui Li and Michael Smidman and Meng Wang and Lin Jiao and Huiqiu Yuan},
  title = {{High-temperature superconductivity with zero-resistance and strange metal behavior in La$_{3}$Ni$_{2}$O$_{7-\delta}$}},
  journal = {Nature Physics},
  volume = {20},
  pages = {1269--1273},
  year = {2024},
  doi = {10.1038/s41567-024-02515-y},
  url = {https://www.nature.com/articles/s41567-024-02515-y}
}

@article{Hou_2023,
doi = {10.1088/0256-307X/40/11/117302},
url = {https://dx.doi.org/10.1088/0256-307X/40/11/117302},
year = {2023},
month = {oct},
publisher = {Chinese Physical Society and IOP Publishing Ltd},
volume = {40},
number = {11},
pages = {117302},
author = {Jun Hou and Peng-Tao Yang and Zi-Yi Liu and Jing-Yuan Li and Peng-Fei Shan and Liang Ma and Gang Wang and Ning-Ning Wang and Hai-Zhong Guo and Jian-Ping Sun and Yoshiya Uwatoko and Meng Wang and Guang-Ming Zhang and Bo-Sen Wang and Jin-Guang Cheng},
title = {{Emergence of High-Temperature Superconducting Phase in Pressurized La$_3$Ni$_2$O$_7$ Crystals}},
journal = {Chinese Physics Letters}
}

@article{Li_2025,
   title={{Angle-resolved photoemission spectroscopy of superconducting (La,Pr)$_3$Ni$_2$O$_7$/SrLaAlO4 heterostructures}},
   ISSN={2053-714X},
   url={http://dx.doi.org/10.1093/nsr/nwaf205},
   DOI={10.1093/nsr/nwaf205},
   journal={National Science Review},
   publisher={Oxford University Press (OUP)},
   author={Li, Peng and Zhou, Guangdi and Lv, Wei and Li, Yueying and Yue, Changming and Huang, Haoliang and Xu, Lizhi and Shen, Jianchang and Miao, Yu and Song, Wenhua and Nie, Zihao and Chen, Yaqi and Wang, Heng and Chen, Weiqiang and Huang, Yaobo and Chen, Zhen-Hua and Qian, Tian and Lin, Junhao and He, Junfeng and Sun, Yu-Jie and Chen, Zhuoyu and Xue, Qi-Kun},
   year={2025},
   month=may }

@misc{wang2025electronicstructurecompressivelystrained,
      title={{Electronic structure of compressively strained thin film La$_2$PrNi$_2$O$_7$}}, 
      author={Bai Yang Wang and Yong Zhong and Sebastien Abadi and Yidi Liu and Yijun Yu and Xiaoliang Zhang and Yi-Ming Wu and Ruohan Wang and Jiarui Li and Yaoju Tarn and Eun Kyo Ko and Vivek Thampy and Makoto Hashimoto and Donghui Lu and Young S. Lee and Thomas P. Devereaux and Chunjing Jia and Harold Y. Hwang and Zhi-Xun Shen},
      year={2025},
      eprint={2504.16372},
      archivePrefix={arXiv},
      primaryClass={cond-mat.supr-con},
      url={https://arxiv.org/abs/2504.16372}, 
}

@article{PhysRevLett.132.256503,
  title = {{Evidence of Spin Density Waves in ${\mathrm{La}}_{3}{\mathrm{Ni}}_{2}{\mathrm{O}}_{7\ensuremath{-}\ensuremath{\delta}}$}},
  author = {Chen, Kaiwen and Liu, Xiangqi and Jiao, Jiachen and Zou, Muyuan and Jiang, Chengyu and Li, Xin and Luo, Yixuan and Wu, Qiong and Zhang, Ningyuan and Guo, Yanfeng and Shu, Lei},
  journal = {Phys. Rev. Lett.},
  volume = {132},
  issue = {25},
  pages = {256503},
  numpages = {7},
  year = {2024},
  month = {Jun},
  publisher = {American Physical Society},
  doi = {10.1103/PhysRevLett.132.256503},
  url = {https://link.aps.org/doi/10.1103/PhysRevLett.132.256503}
}

@Article{Chen2024,
author={Chen, Xiaoyang
and Choi, Jaewon
and Jiang, Zhicheng
and Mei, Jiong
and Jiang, Kun
and Li, Jie
and Agrestini, Stefano
and Garcia-Fernandez, Mirian
and Sun, Hualei
and Huang, Xing
and Shen, Dawei
and Wang, Meng
and Hu, Jiangping
and Lu, Yi
and Zhou, Ke-Jin
and Feng, Donglai},
title={{Electronic and magnetic excitations in La$_3$Ni$_2$O$_7$}},
journal={Nature Communications},
year={2024},
month={Nov},
day={06},
volume={15},
number={1},
pages={9597},
issn={2041-1723},
doi={10.1038/s41467-024-53863-5},
url={https://doi.org/10.1038/s41467-024-53863-5}
}

@Article{Zhang2024,
author={Zhang, Yang
and Lin, Ling-Fang
and Moreo, Adriana
and Maier, Thomas A.
and Dagotto, Elbio},
title={{Structural phase transition, s{\textpm}-wave pairing, and magnetic stripe order in bilayered superconductor La$_3$Ni$_2$O$_7$ under pressure}},
journal={Nature Communications},
year={2024},
month={Mar},
day={19},
volume={15},
number={1},
pages={2470},
issn={2041-1723},
doi={10.1038/s41467-024-46622-z},
url={https://doi.org/10.1038/s41467-024-46622-z}
}

@misc{zhan2024cooperation,
      title={Cooperation between electron-phonon coupling and electronic interaction in bilayer nickelates La$_3$Ni$_2$O$_7$}, 
      author={Jun Zhan and Yuhao Gu and Xianxin Wu and Jiangping Hu},
      year={2024},
      eprint={2404.03638},
      archivePrefix={arXiv},
      primaryClass={cond-mat.supr-con}
}

@article{PhysRevMaterials.8.074802,
  title = {{Electronic instability, layer selectivity, and Fermi arcs in ${\text{La}}_{3}{\text{Ni}}_{2}{\text{O}}_{7}$}},
  author = {Lechermann, Frank and B\"otzel, Steffen and Eremin, Ilya M.},
  journal = {Phys. Rev. Mater.},
  volume = {8},
  issue = {7},
  pages = {074802},
  numpages = {7},
  year = {2024},
  month = {Jul},
  publisher = {American Physical Society},
  doi = {10.1103/PhysRevMaterials.8.074802},
  url = {https://link.aps.org/doi/10.1103/PhysRevMaterials.8.074802}
}

@article{wang2024pressure,
  title={{Pressure-induced superconductivity in polycrystalline La$_3$Ni$_2$O$_{7-\delta}$}},
  author={Wang, G and Wang, NN and Shen, XL and Hou, J and Ma, L and Shi, LF and Ren, ZA and Gu, YD and Ma, HM and Yang, PT and others},
  journal={Physical Review X},
  volume={14},
  number={1},
  pages={011040},
  year={2024},
  publisher={APS}
}

@article{Xue_2024,
doi = {10.1088/0256-307X/41/5/057403},
url = {https://dx.doi.org/10.1088/0256-307X/41/5/057403},
year = {2024},
month = {may},
publisher = {Chinese Physical Society and IOP Publishing Ltd},
volume = {41},
number = {5},
pages = {057403},
author = {Xue, Jie-Ran and Wang, Fa},
title = {{Magnetism and Superconductivity in the t–J Model of La$_3$Ni$_2$O$_7$ Under Multiband Gutzwiller Approximation}},
journal = {Chinese Physics Letters}
}

@article{jiang2024high,
  title={{High-temperature superconductivity in La$_3$Ni$_2$O$_7$}},
  author={Jiang, Kun and Wang, Ziqiang and Zhang, Fu-Chun},
  journal={Chinese Physics Letters},
  volume={41},
  number={1},
  pages={017402},
  year={2024},
  publisher={IOP Publishing}
}

@article{PhysRevB.108.214522,
  title = {{Electron correlations and superconductivity in ${\mathrm{La}}_{3}{\mathrm{Ni}}_{2}{\mathrm{O}}_{7}$ under pressure tuning}},
  author = {Liao, Zhiguang and Chen, Lei and Duan, Guijing and Wang, Yiming and Liu, Changle and Yu, Rong and Si, Qimiao},
  journal = {Phys. Rev. B},
  volume = {108},
  issue = {21},
  pages = {214522},
  numpages = {9},
  year = {2023},
  month = {Dec},
  publisher = {American Physical Society},
  doi = {10.1103/PhysRevB.108.214522},
  url = {https://link.aps.org/doi/10.1103/PhysRevB.108.214522}
}

@article{fabian1,
  title = {Pairing dome from an emergent Feshbach resonance in a strongly repulsive bilayer model},
  author = {Lange, Hannah and Homeier, Lukas and Demler, Eugene and Schollw\"ock, Ulrich and Bohrdt, Annabelle and Grusdt, Fabian},
  journal = {Phys. Rev. B},
  volume = {110},
  issue = {8},
  pages = {L081113},
  numpages = {7},
  year = {2024},
  month = {Aug},
  publisher = {American Physical Society},
  doi = {10.1103/PhysRevB.110.L081113},
  url = {https://link.aps.org/doi/10.1103/PhysRevB.110.L081113}
}

@misc{fabian2,
      title={Feshbach resonance in a strongly repulsive bilayer model: a possible scenario for bilayer nickelate superconductors}, 
      author={Hannah Lange and Lukas Homeier and Eugene Demler and Ulrich Schollwöck and Fabian Grusdt and Annabelle Bohrdt},
      year={2023},
      eprint={2309.15843},
      archivePrefix={arXiv},
      primaryClass={cond-mat.str-el}
}

@misc{fabian3,
      title={{Superconductivity in the pressurized nickelate La$_3$Ni$_2$O$_7$ in the vicinity of a BEC-BCS crossover}}, 
      author={Henning Schlömer and Ulrich Schollwöck and Fabian Grusdt and Annabelle Bohrdt},
      year={2023},
      eprint={2311.03349},
      archivePrefix={arXiv},
      primaryClass={cond-mat.str-el}
}

@article{PhysRevLett.132.146002,
  title = {{Interlayer-Coupling-Driven High-Temperature Superconductivity in ${\mathrm{La}}_{3}{\mathrm{Ni}}_{2}{\mathrm{O}}_{7}$ under Pressure}},
  author = {Lu, Chen and Pan, Zhiming and Yang, Fan and Wu, Congjun},
  journal = {Phys. Rev. Lett.},
  volume = {132},
  issue = {14},
  pages = {146002},
  numpages = {6},
  year = {2024},
  month = {Apr},
  publisher = {American Physical Society},
  doi = {10.1103/PhysRevLett.132.146002},
  url = {https://link.aps.org/doi/10.1103/PhysRevLett.132.146002}
}

@article{PhysRevB.110.104517,
  title = {{Strong pairing from a small Fermi surface beyond weak coupling: Application to ${\mathrm{La}}_{3}{\mathrm{Ni}}_{2}{\mathrm{O}}_{7}$}},
  author = {Yang, Hui and Oh, Hanbit and Zhang, Ya-Hui},
  journal = {Phys. Rev. B},
  volume = {110},
  issue = {10},
  pages = {104517},
  numpages = {21},
  year = {2024},
  month = {Sep},
  publisher = {American Physical Society},
  doi = {10.1103/PhysRevB.110.104517},
  url = {https://link.aps.org/doi/10.1103/PhysRevB.110.104517}
}

@article{PhysRevB.109.L081105,
  title = {{Flat bands promoted by Hund's rule coupling in the candidate double-layer high-temperature superconductor ${\mathrm{La}}_{3}{\mathrm{Ni}}_{2}{\mathrm{O}}_{7}$ under high pressure}},
  author = {Cao, Yingying and Yang, Yi-feng},
  journal = {Phys. Rev. B},
  volume = {109},
  issue = {8},
  pages = {L081105},
  numpages = {6},
  year = {2024},
  month = {Feb},
  publisher = {American Physical Society},
  doi = {10.1103/PhysRevB.109.L081105},
  url = {https://link.aps.org/doi/10.1103/PhysRevB.109.L081105}
}

@article{PhysRevB.111.174506,
  title = {{Effective model and pairing tendency in the bilayer Ni-based superconductor ${\mathrm{La}}_{3}{\mathrm{Ni}}_{2}{\mathrm{O}}_{7}$}},
  author = {Gu, Yuhao and Le, Congcong and Yang, Zhesen and Wu, Xianxin and Hu, Jiangping},
  journal = {Phys. Rev. B},
  volume = {111},
  issue = {17},
  pages = {174506},
  numpages = {7},
  year = {2025},
  month = {May},
  publisher = {American Physical Society},
  doi = {10.1103/PhysRevB.111.174506},
  url = {https://link.aps.org/doi/10.1103/PhysRevB.111.174506}
}

@article{PhysRevB.111.L020504,
  title = {{Type-II $t\text{\ensuremath{-}}J$ model in charge transfer regime in bilayer ${\mathrm{La}}_{3}{\mathrm{Ni}}_{2}{\mathrm{O}}_{7}$ and trilayer ${\mathrm{La}}_{4}{\mathrm{Ni}}_{3}{\mathrm{O}}_{10}$}},
  author = {Oh, Hanbit and Zhou, Boran and Zhang, Ya-Hui},
  journal = {Phys. Rev. B},
  volume = {111},
  issue = {2},
  pages = {L020504},
  numpages = {7},
  year = {2025},
  month = {Jan},
  publisher = {American Physical Society},
  doi = {10.1103/PhysRevB.111.L020504},
  url = {https://link.aps.org/doi/10.1103/PhysRevB.111.L020504}
}

@article{PhysRevLett.131.236002,
  title = {{${s}^{\ifmmode\pm\else\textpm\fi{}}$-Wave Pairing and the Destructive Role of Apical-Oxygen Deficiencies in ${\mathrm{La}}_{3}{\mathrm{Ni}}_{2}{\mathrm{O}}_{7}$ under Pressure}},
  author = {Liu, Yu-Bo and Mei, Jia-Wei and Ye, Fei and Chen, Wei-Qiang and Yang, Fan},
  journal = {Phys. Rev. Lett.},
  volume = {131},
  issue = {23},
  pages = {236002},
  numpages = {6},
  year = {2023},
  month = {Dec},
  publisher = {American Physical Society},
  doi = {10.1103/PhysRevLett.131.236002},
  url = {https://link.aps.org/doi/10.1103/PhysRevLett.131.236002}
}

@article{PhysRevLett.131.206501,
  title = {{Correlated Electronic Structure of ${\mathrm{La}}_{3}{\text{Ni}}_{2}{\mathrm{O}}_{7}$ under Pressure}},
  author = {Christiansson, Viktor and Petocchi, Francesco and Werner, Philipp},
  journal = {Phys. Rev. Lett.},
  volume = {131},
  issue = {20},
  pages = {206501},
  numpages = {6},
  year = {2023},
  month = {Nov},
  publisher = {American Physical Society},
  doi = {10.1103/PhysRevLett.131.206501},
  url = {https://link.aps.org/doi/10.1103/PhysRevLett.131.206501}
}

@Article{Liu2024,
author={Liu, Zhe
and Huo, Mengwu
and Li, Jie
and Li, Qing
and Liu, Yuecong
and Dai, Yaomin
and Zhou, Xiaoxiang
and Hao, Jiahao
and Lu, Yi
and Wang, Meng
and Wen, Hai-Hu},
title={{Electronic correlations and partial gap in the bilayer nickelate La$_3$Ni$_2$O$_7$}},
journal={Nature Communications},
year={2024},
month={Aug},
day={31},
volume={15},
number={1},
pages={7570},
issn={2041-1723},
doi={10.1038/s41467-024-52001-5},
url={https://doi.org/10.1038/s41467-024-52001-5}
}

@article{PhysRevLett.132.126503,
  title = {{Pressure Driven Fractionalization of Ionic Spins Results in Cupratelike High-${T}_{c}$ Superconductivity in ${\mathrm{La}}_{3}{\mathrm{Ni}}_{2}{\mathrm{O}}_{7}$}},
  author = {Jiang, Ruoshi and Hou, Jinning and Fan, Zhiyu and Lang, Zi-Jian and Ku, Wei},
  journal = {Phys. Rev. Lett.},
  volume = {132},
  issue = {12},
  pages = {126503},
  numpages = {7},
  year = {2024},
  month = {Mar},
  publisher = {American Physical Society},
  doi = {10.1103/PhysRevLett.132.126503},
  url = {https://link.aps.org/doi/10.1103/PhysRevLett.132.126503}
}

@article{PhysRevLett.132.036502,
  title = {{Bilayer ${t\text{\ensuremath{-}}J\text{\ensuremath{-}}J}_{\ensuremath{\perp}}$ Model and Magnetically Mediated Pairing in the Pressurized Nickelate ${\mathrm{La}}_{3}{\mathrm{Ni}}_{2}{\mathrm{O}}_{7}$}},
  author = {Qu, Xing-Zhou and Qu, Dai-Wei and Chen, Jialin and Wu, Congjun and Yang, Fan and Li, Wei and Su, Gang},
  journal = {Phys. Rev. Lett.},
  volume = {132},
  issue = {3},
  pages = {036502},
  numpages = {6},
  year = {2024},
  month = {Jan},
  publisher = {American Physical Society},
  doi = {10.1103/PhysRevLett.132.036502},
  url = {https://link.aps.org/doi/10.1103/PhysRevLett.132.036502}
}

@Article{Yang2024,
author={Yang, Jiangang
and Sun, Hualei
and Hu, Xunwu
and Xie, Yuyang
and Miao, Taimin
and Luo, Hailan
and Chen, Hao
and Liang, Bo
and Zhu, Wenpei
and Qu, Gexing
and Chen, Cui-Qun
and Huo, Mengwu
and Huang, Yaobo
and Zhang, Shenjin
and Zhang, Fengfeng
and Yang, Feng
and Wang, Zhimin
and Peng, Qinjun
and Mao, Hanqing
and Liu, Guodong
and Xu, Zuyan
and Qian, Tian
and Yao, Dao-Xin
and Wang, Meng
and Zhao, Lin
and Zhou, X. J.},
title={{Orbital-dependent electron correlation in double-layer nickelate La$_3$Ni$_2$O$_7$}},
journal={Nature Communications},
year={2024},
month={May},
day={23},
volume={15},
number={1},
pages={4373},
issn={2041-1723},
doi={10.1038/s41467-024-48701-7},
url={https://doi.org/10.1038/s41467-024-48701-7}
}

@article{ZHANG2024147,
title = {{Effects of pressure and doping on Ruddlesden-Popper phases La$_{n+1}$Ni$_n$O$_{3n+1}$}},
journal = {Journal of Materials Science $\&$ Technology},
volume = {185},
pages = {147-154},
year = {2024},
issn = {1005-0302},
doi = {https://doi.org/10.1016/j.jmst.2023.11.011},
url = {https://www.sciencedirect.com/science/article/pii/S1005030223009829},
author = {Mingxin Zhang and Cuiying Pei and Qi Wang and Yi Zhao and Changhua Li and Weizheng Cao and Shihao Zhu and Juefei Wu and Yanpeng Qi}
}

@article{lu2023superconductivity,
  title={{Superconductivity from Doping Symmetric Mass Generation Insulators: Application to La$_3$Ni$_2$O$_7$ under Pressure}},
  author={Lu, Da-Chuan and Li, Miao and Zeng, Zhao-Yi and Hou, Wanda and Wang, Juven and Yang, Fan and You, Yi-Zhuang},
  journal={arXiv preprint arXiv:2308.11195},
  year={2023}
}

@article{PhysRevB.108.174501,
  title = {{Impurity and vortex states in the bilayer high-temperature superconductor ${\mathrm{La}}_{3}{\mathrm{Ni}}_{2}{\mathrm{O}}_{7}$}},
  author = {Huang, Junkang and Wang, Z. D. and Zhou, Tao},
  journal = {Phys. Rev. B},
  volume = {108},
  issue = {17},
  pages = {174501},
  numpages = {7},
  year = {2023},
  month = {Nov},
  publisher = {American Physical Society},
  doi = {10.1103/PhysRevB.108.174501},
  url = {https://link.aps.org/doi/10.1103/PhysRevB.108.174501}
}

@article{PhysRevB.108.165141,
  title = {{Trends in electronic structures and ${s}_{\ifmmode\pm\else\textpm\fi{}}$-wave pairing for the rare-earth series in bilayer nickelate superconductor ${R}_{3}{\mathrm{Ni}}_{2}{\mathrm{O}}_{7}$}},
  author = {Zhang, Yang and Lin, Ling-Fang and Moreo, Adriana and Maier, Thomas A. and Dagotto, Elbio},
  journal = {Phys. Rev. B},
  volume = {108},
  issue = {16},
  pages = {165141},
  numpages = {8},
  year = {2023},
  month = {Oct},
  publisher = {American Physical Society},
  doi = {10.1103/PhysRevB.108.165141},
  url = {https://link.aps.org/doi/10.1103/PhysRevB.108.165141}
}

@article{PhysRevB.108.L201108,
  title = {{Interlayer valence bonds and two-component theory for high-${T}_{c}$ superconductivity of ${\mathrm{La}}_{3}{\mathrm{Ni}}_{2}{\mathrm{O}}_{7}$ under pressure}},
  author = {Yang, Yi-feng and Zhang, Guang-Ming and Zhang, Fu-Chun},
  journal = {Phys. Rev. B},
  volume = {108},
  issue = {20},
  pages = {L201108},
  numpages = {6},
  year = {2023},
  month = {Nov},
  publisher = {American Physical Society},
  doi = {10.1103/PhysRevB.108.L201108},
  url = {https://link.aps.org/doi/10.1103/PhysRevB.108.L201108}
}

@ARTICLE{2023arXiv230812750K,
       author = {{Kitamine}, Naoya and {Ochi}, Masayuki and {Kuroki}, Kazuhiko},
        title = {{Theoretical designing of multiband Nickelate and Palladate superconductors with $d^{8+\delta}$ configuration}},
      journal = {arXiv e-prints},
         year = 2023,
        month = aug,
          eid = {arXiv:2308.12750},
        pages = {arXiv:2308.12750},
          doi = {10.48550/arXiv.2308.12750},
archivePrefix = {arXiv},
       eprint = {2308.12750},
 primaryClass = {cond-mat.supr-con},
       adsurl = {https://ui.adsabs.harvard.edu/abs/2023arXiv230812750K}
}

@article{PhysRevB.108.L140505,
  title = {{Possible ${s}_{\ifmmode\pm\else\textpm\fi{}}$-wave superconductivity in ${\mathrm{La}}_{3}{\mathrm{Ni}}_{2}{\mathrm{O}}_{7}$}},
  author = {Yang, Qing-Geng and Wang, Da and Wang, Qiang-Hua},
  journal = {Phys. Rev. B},
  volume = {108},
  issue = {14},
  pages = {L140505},
  numpages = {5},
  year = {2023},
  month = {Oct},
  publisher = {American Physical Society},
  doi = {10.1103/PhysRevB.108.L140505},
  url = {https://link.aps.org/doi/10.1103/PhysRevB.108.L140505}
}

@article{shen2023effective,
  title={{Effective bi-layer model Hamiltonian and density-matrix renormalization group study for the high-T c superconductivity in La$_3$Ni$_2$O$_7$ under high pressure}},
  author={Shen, Yang and Qin, Mingpu and Zhang, Guang-Ming},
  journal={Chinese Physics Letters},
  volume={40},
  number={12},
  pages={127401},
  year={2023},
  publisher={IOP Publishing}
}

@article{PhysRevLett.133.126501,
  title = {{Strong Pairing Originated from an Emergent ${\mathbb{Z}}_{2}$ Berry Phase in ${\mathrm{La}}_{3}{\mathrm{Ni}}_{2}{\mathrm{O}}_{7}$}},
  author = {Zhang, Jia-Xin and Zhang, Hao-Kai and You, Yi-Zhuang and Weng, Zheng-Yu},
  journal = {Phys. Rev. Lett.},
  volume = {133},
  issue = {12},
  pages = {126501},
  numpages = {7},
  year = {2024},
  month = {Sep},
  publisher = {American Physical Society},
  doi = {10.1103/PhysRevLett.133.126501},
  url = {https://link.aps.org/doi/10.1103/PhysRevLett.133.126501}
}

@article{PhysRevB.108.L140504,
  title = {{High-${T}_{c}$ superconductivity by mobilizing local spin singlets and possible route to higher ${T}_{c}$ in pressurized ${\mathrm{La}}_{3}{\mathrm{Ni}}_{2}{\mathrm{O}}_{7}$}},
  author = {Qin, Qiong and Yang, Yi-feng},
  journal = {Phys. Rev. B},
  volume = {108},
  issue = {14},
  pages = {L140504},
  numpages = {6},
  year = {2023},
  month = {Oct},
  publisher = {American Physical Society},
  doi = {10.1103/PhysRevB.108.L140504},
  url = {https://link.aps.org/doi/10.1103/PhysRevB.108.L140504}
}

@misc{guo2025revealingsuperconductinggap,
      title={Revealing superconducting gap in La$_3$Ni$_2$O$_7$-$\delta$ by Andreev reflection spectroscopy under high pressure}, 
      author={Jianning Guo and Yuzhi Chen and Yulong Wang and Hualei Sun and Deyuan Hu and Meng Wang and Xiaoli Huang and Tian Cui},
      year={2025},
      eprint={2509.12601},
      archivePrefix={arXiv},
      primaryClass={cond-mat.supr-con},
      url={https://arxiv.org/abs/2509.12601}, 
}

@misc{cao2025directobservationdwave,
      title={Direct Observation of d-Wave Superconducting Gap Symmetry in Pressurized La3Ni2O7-delta Single Crystals}, 
      author={Zi-Yu Cao and Di Peng and Seokmin Choi and Fujun Lan and Lan Yu and Enkang Zhang and Zhenfang Xing and Yuxin Liu and Feiyang Zhang and Tao Luo and Lixing Chen and Vuong Thi Anh Hong and Seung-Yeop Paek and Harim Jang and Jinghong Xie and Huayu Liu and Hongbo Lou and Zhidan Zeng and Yang Ding and Jun Zhao and Cailong Liu and Tuson Park and Qiaoshi Zeng and Ho-kwang Mao},
      year={2025},
      eprint={2509.12606},
      archivePrefix={arXiv},
      primaryClass={cond-mat.supr-con},
      url={https://arxiv.org/abs/2509.12606}, 
}

@article{PhysRevLett.133.096002,
  title = {Quenched Pair Breaking by Interlayer Correlations as a Key to Superconductivity in ${\mathrm{La}}_{3}{\mathrm{Ni}}_{2}{\mathrm{O}}_{7}$},
  author = {Ryee, Siheon and Witt, Niklas and Wehling, Tim O.},
  journal = {Phys. Rev. Lett.},
  volume = {133},
  issue = {9},
  pages = {096002},
  numpages = {7},
  year = {2024},
  month = {Aug},
  publisher = {American Physical Society},
  doi = {10.1103/PhysRevLett.133.096002},
  url = {https://link.aps.org/doi/10.1103/PhysRevLett.133.096002}
}

@article{ncbf-9b8m,
  title = {Superconductivity Governed by Janus-Faced Fermiology in Strained Bilayer Nickelates},
  author = {Ryee, Siheon and Witt, Niklas and Sangiovanni, Giorgio and Wehling, Tim O.},
  journal = {Phys. Rev. Lett.},
  volume = {135},
  issue = {23},
  pages = {236003},
  numpages = {8},
  year = {2025},
  month = {Dec},
  publisher = {American Physical Society},
  doi = {10.1103/ncbf-9b8m},
  url = {https://link.aps.org/doi/10.1103/ncbf-9b8m}
}

\end{document}